\documentclass[twocolumn,runningheads]{svjour3}

\usepackage[latin1]{inputenc}
\usepackage{times}

\usepackage{psfrag,amsmath, amssymb, amscd, amsfonts, latexsym, epsf, graphicx,amscd}
\usepackage{natbib}
\usepackage[switch]{lineno}

\def\bbbr{{\mathbb R}} 
 
\def\diag{\operatorname{diag}}

\def\norm{\scriptsize\mbox{norm}}
\def\simple{\scriptsize\mbox{simple}}

\def\spat{\scriptsize\mbox{spat}}
\def\vel{\scriptsize\mbox{vel}}
\def\even{\scriptsize\mbox{even}}
\def\odd{\scriptsize\mbox{odd}}
\def\complex{\scriptsize\mbox{complex}}

\journalname{arXiv preprint}

\begin{document}

\titlerunning{Do the receptive fields in the primary visual cortex span a variability over the degree of elongation of the receptive fields?}

\title{\bf Do the receptive fields in the primary visual cortex span a variability over the degree of elongation of the receptive fields?%
\thanks{The support from the Swedish Research Council 
              (contract 2022-02969) is gratefully acknowledged. }}

\author{Tony Lindeberg}

\institute{Computational Brain Science Lab,
        Division of Computational Science and Technology,
        KTH Royal Institute of Technology,
        SE-100 44 Stockholm, Sweden.
       \email{tony@kth.se}. ORCID: 0000-0002-9081-2170.}

\date{}

\maketitle

\begin{abstract}
  \noindent
  This paper presents the results of combining (i) theoretical
  analysis regarding connections between the orientation selectivity
  and the elongation of receptive fields for the affine Gaussian
  derivative model with (ii) biological measurements of orientation
  selectivity in the primary visual cortex to investigate if (iii) the
  receptive fields can be regarded as spanning a variability in the
  degree of elongation.

From an in-depth theoretical analysis of idealized models for the
receptive fields of simple and complex cells in the primary visual
cortex, we established that the orientation selectivity becomes
more narrow with increasing elongation of the receptive
fields. Combined with previously established biological results,
concerning broad vs. sharp orientation tuning of visual neurons in the
primary visual cortex, as well as previous experimental results
concerning distributions of the resultant of the orientation
selectivity curves for simple and complex cells,
we show that these results are consistent with
the receptive fields spanning a variability over the degree of elongation of the receptive fields.
We also show that our principled theoretical model for visual
receptive fields leads to qualitatively similar types of deviations
from a uniform histogram of the resultant descriptor of the orientation
selectivity curves for simple cells, as can be observed in the results from
biological experiments.

To firmly investigate the validity of the underlying working hypothesis,
we finally formulate a set of testable predictions for biological
experiments, to characterize the predicted systematic variability
in the elongation over the orientation maps in higher mammals, and
its relations to the pinwheel structure.


\end{abstract}

\begin{figure*}[hbtp]
\begin{center}
    \begin{tabular}{cccc}
      \includegraphics[width=0.30\textwidth]{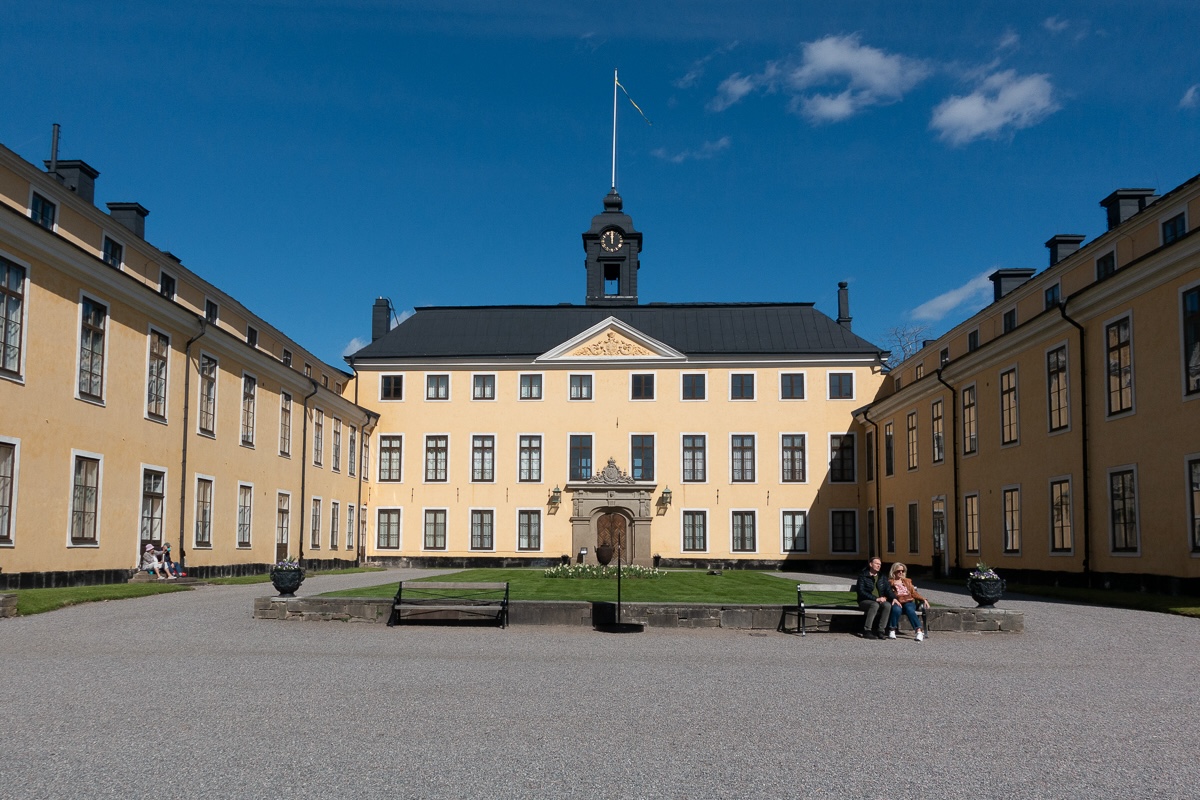}
       & \includegraphics[width=0.30\textwidth]{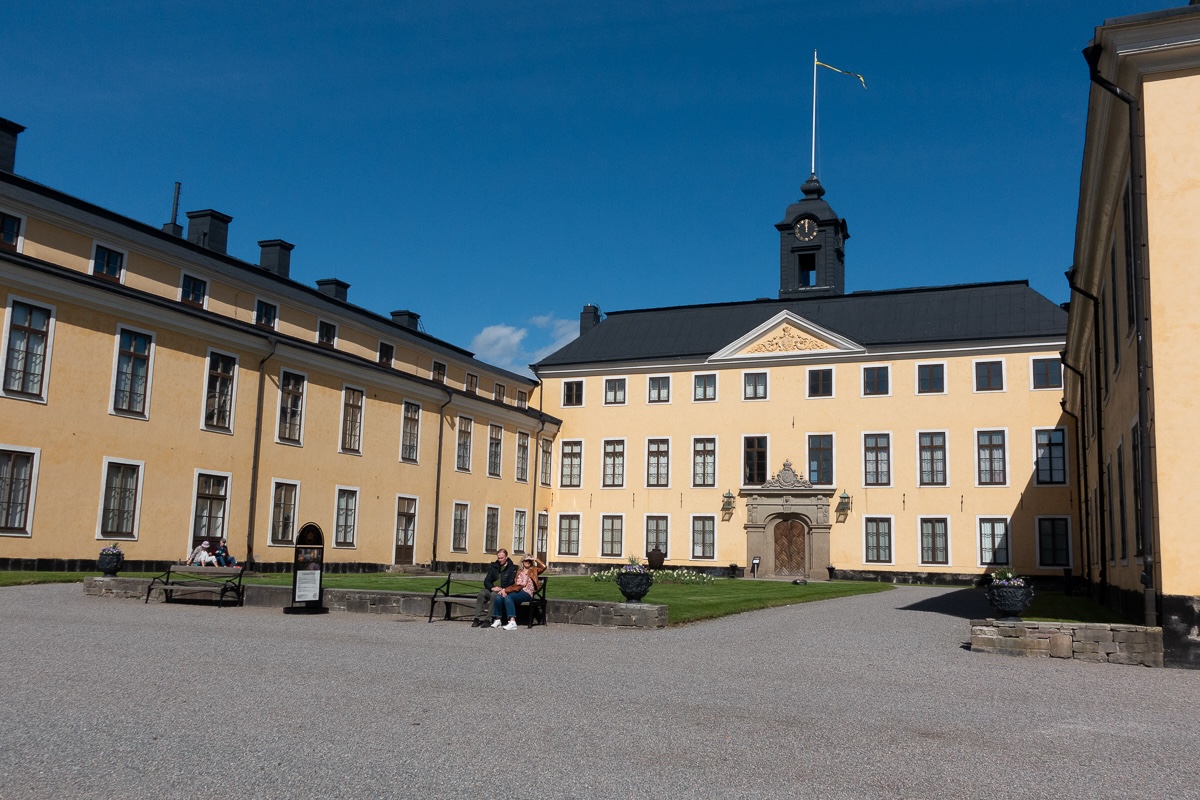}      
       & \includegraphics[width=0.30\textwidth]{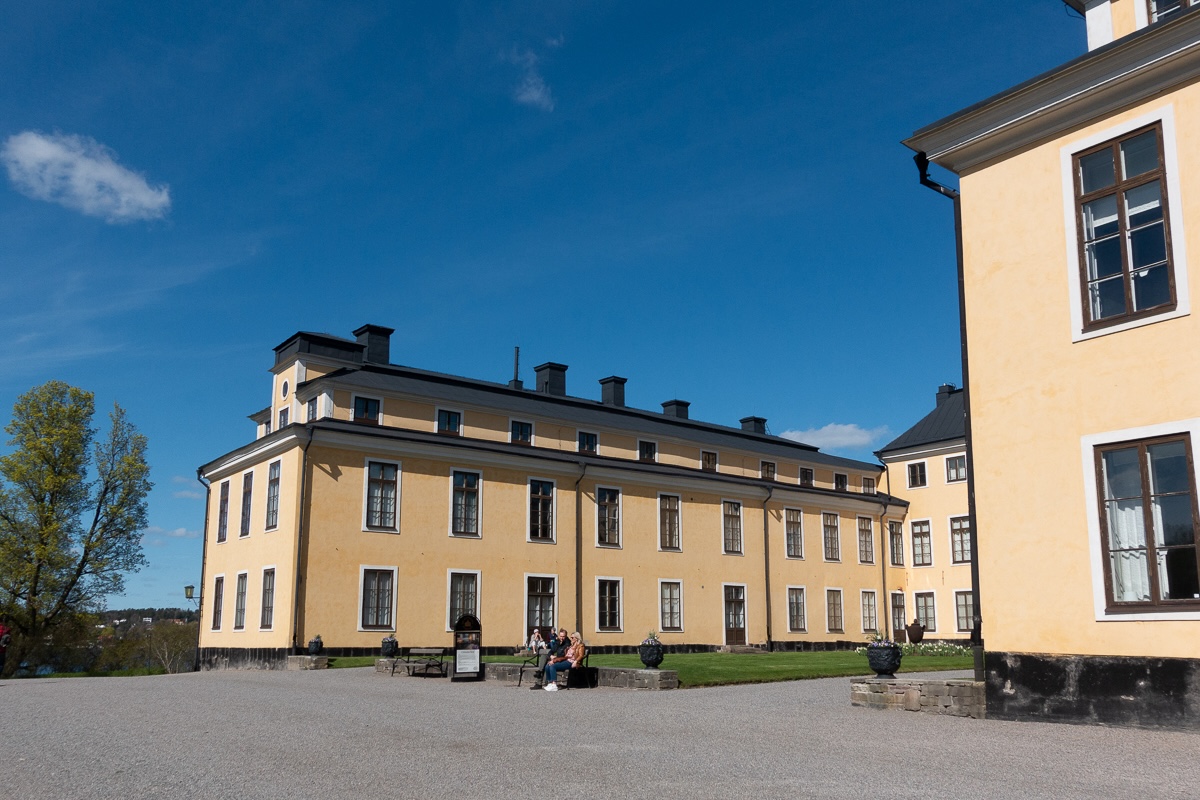}
    \end{tabular}
  \end{center}
  \caption{Variabilities in image structures generated by viewing the
    same surface patterns from different viewing directions. Observe
    how the resulting perspective transformations lead to strong 
    foreshortening effects, in that the image structures in one
    direction in the 2-D image space are compressed more than
    the image structures in the orthogonal direction. Here, where the
    viewpoint of the observer is moved
    horizontally in the world, the foreshortening effect is mainly along the horizontal
    direction, although complemented also with small rotations for
    non-central image points, because of using a planar image plane
    as opposed to a spherical retina. To first order of approximation of the projective
    mappings between pairwise views, these resulting image deformations
    can be modelled in terms of local affine transformations.}
  \label{fig-castle-view-dirs}
\end{figure*}

\begin{figure*}[btp]
\begin{center}
    \begin{tabular}{cccccc}
      $\kappa= 1$
      & $\kappa = 2 \sqrt{2}$
      & $\kappa = 2$
      & $\kappa = 2\sqrt{2}$
      & $\kappa = 4$        
      & $\kappa = 4\sqrt{2}$ \\
     \includegraphics[width=0.14\textwidth]{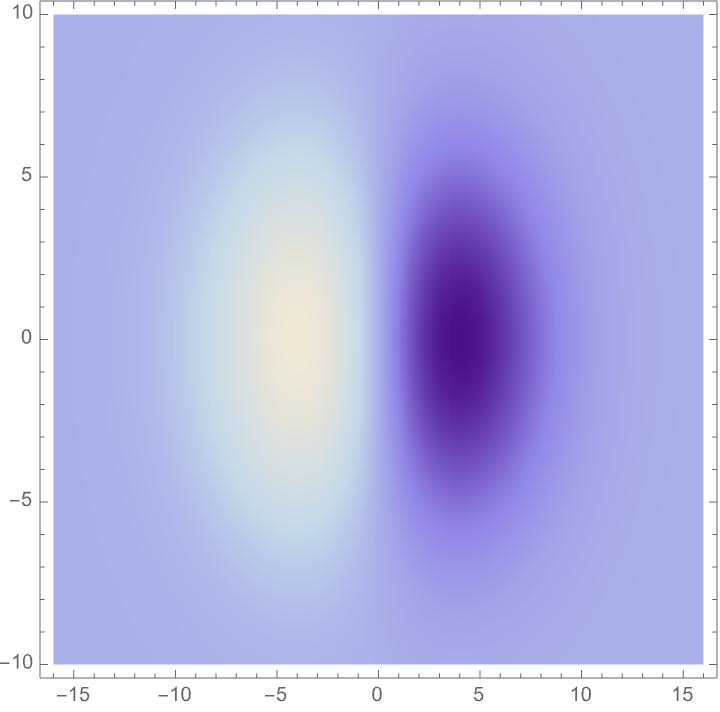}
      & \includegraphics[width=0.14\textwidth]{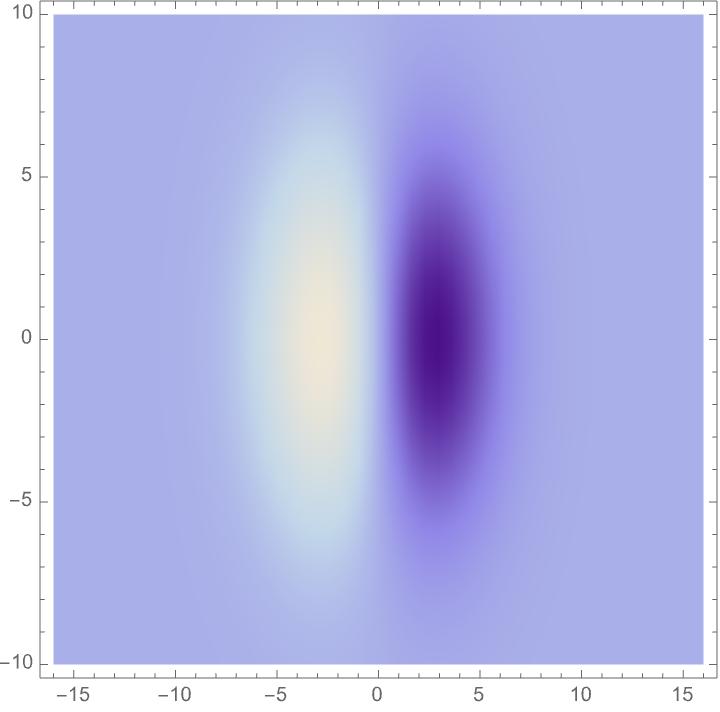}
      & \includegraphics[width=0.14\textwidth]{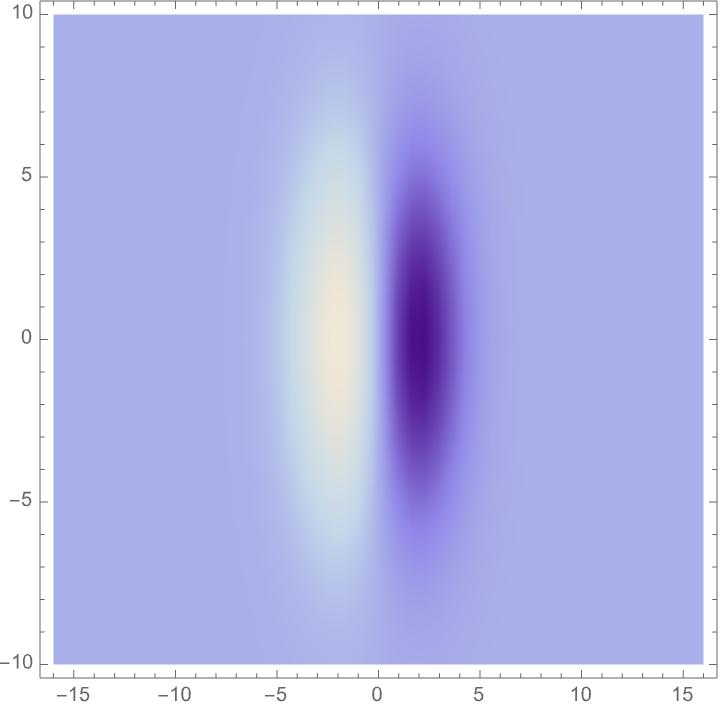}
      & \includegraphics[width=0.14\textwidth]{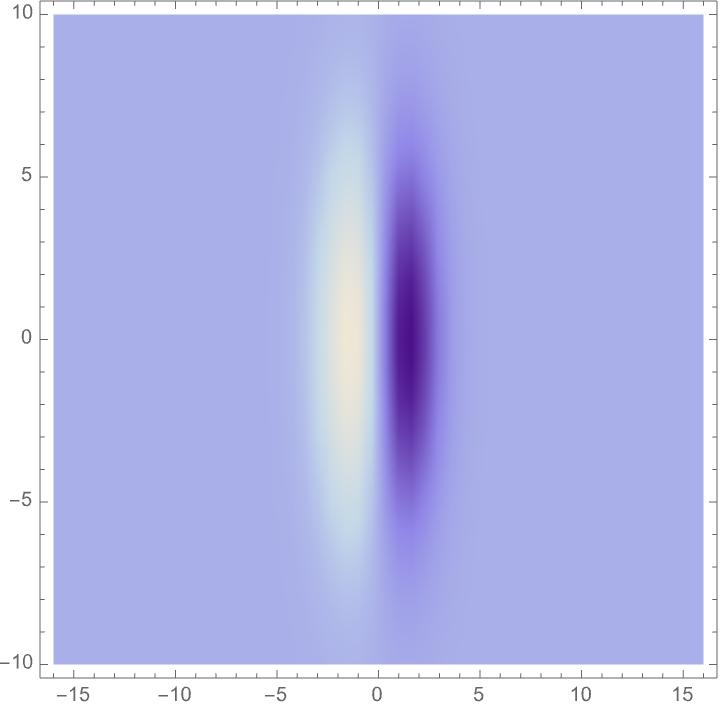}
      & \includegraphics[width=0.14\textwidth]{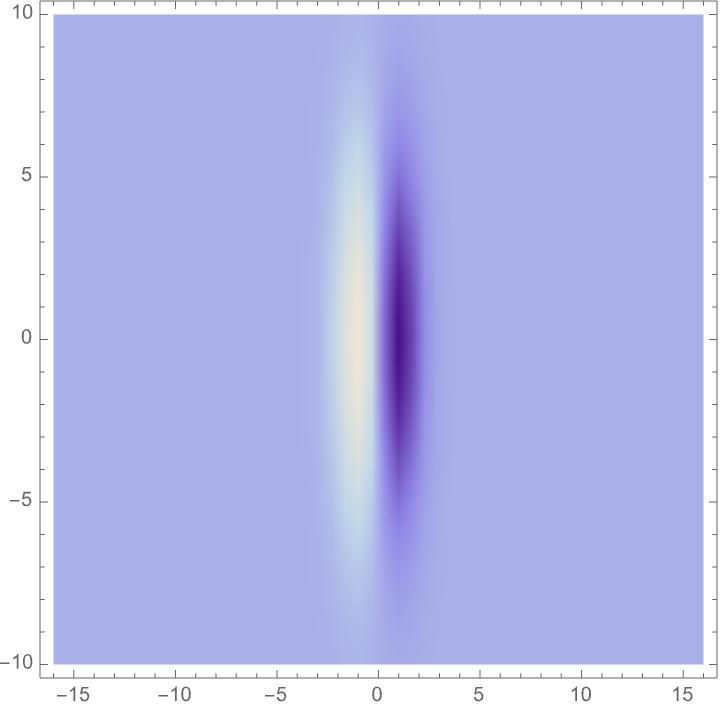}
      & \includegraphics[width=0.14\textwidth]{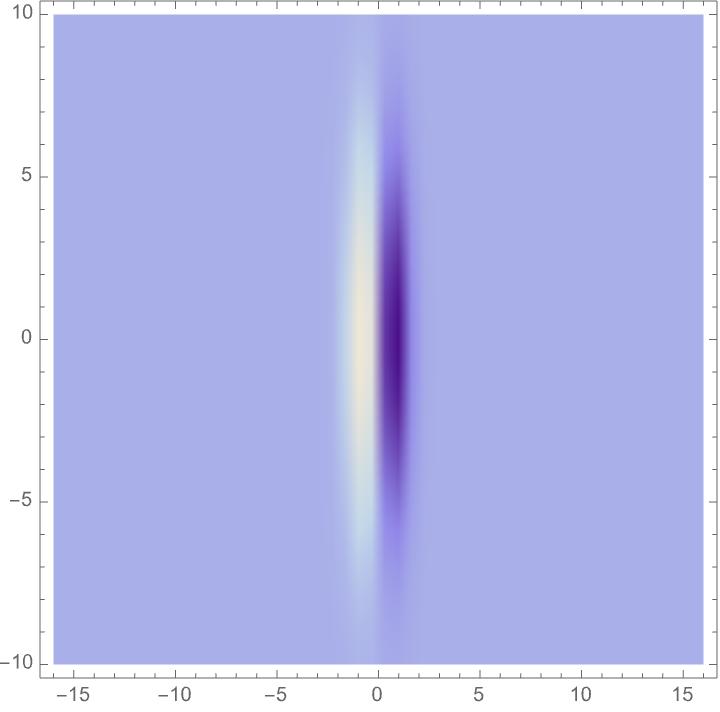} \\
     \includegraphics[width=0.14\textwidth]{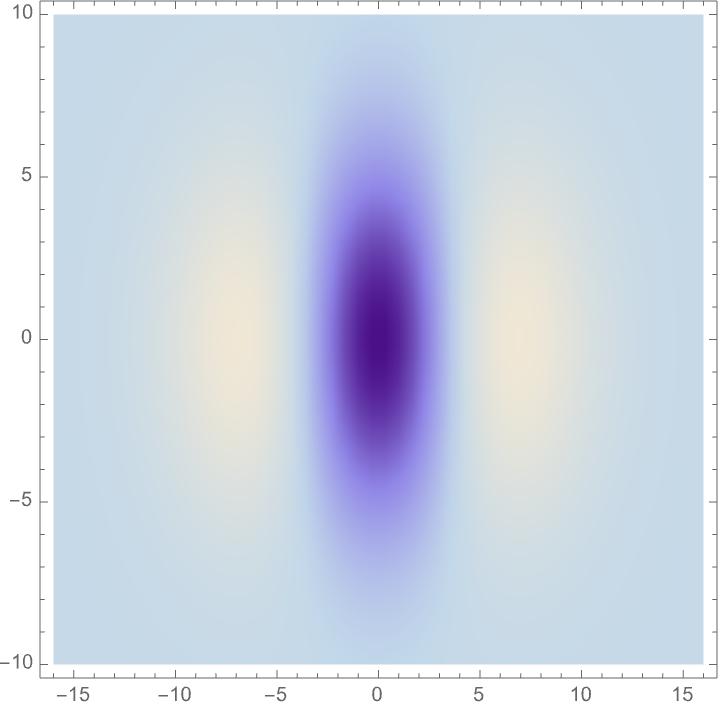}
      & \includegraphics[width=0.14\textwidth]{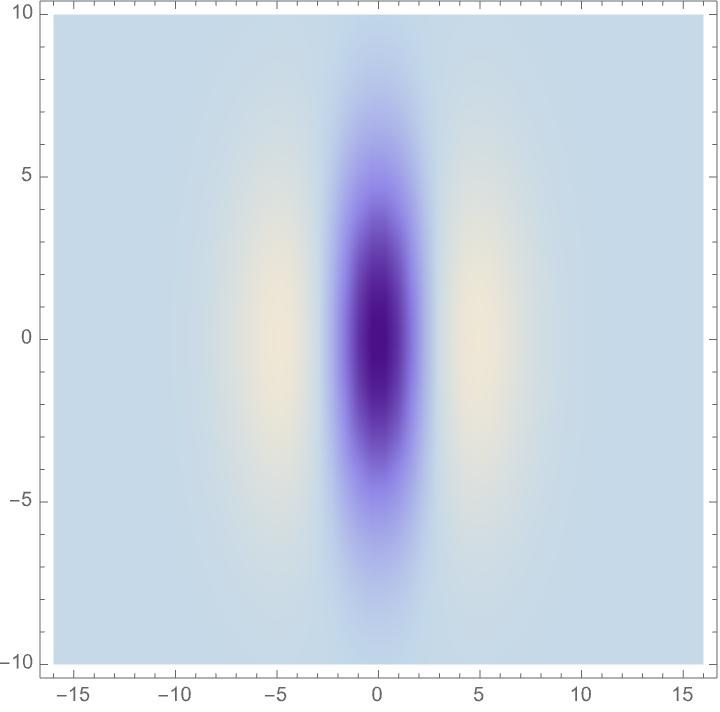}
      & \includegraphics[width=0.14\textwidth]{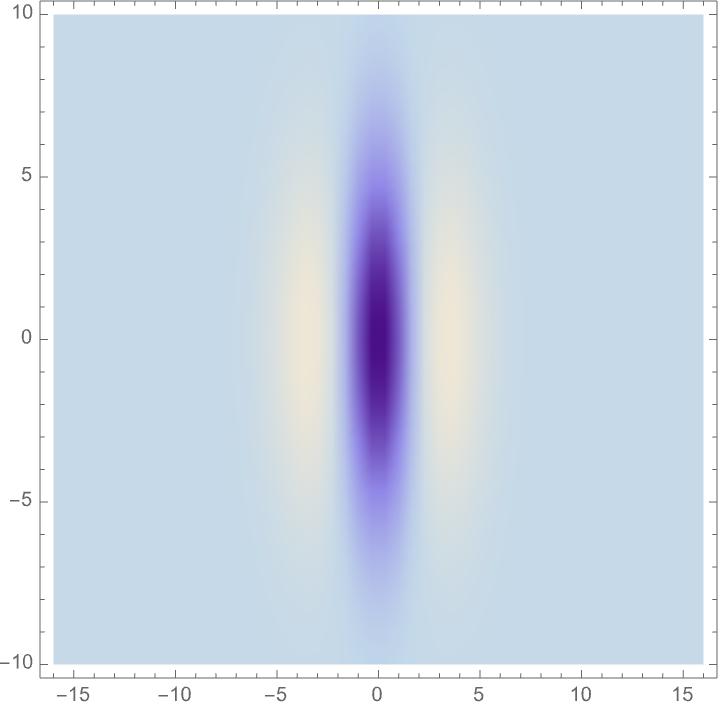}
      & \includegraphics[width=0.14\textwidth]{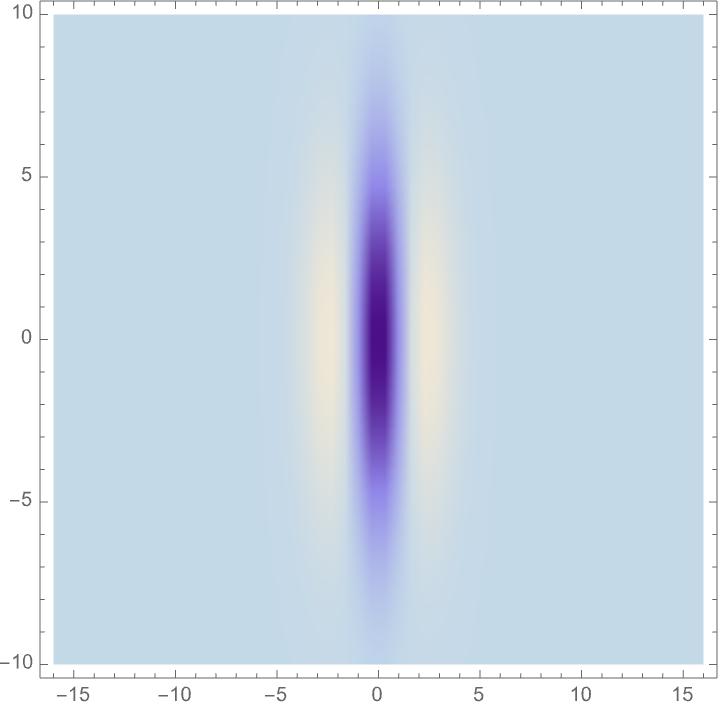}
      & \includegraphics[width=0.14\textwidth]{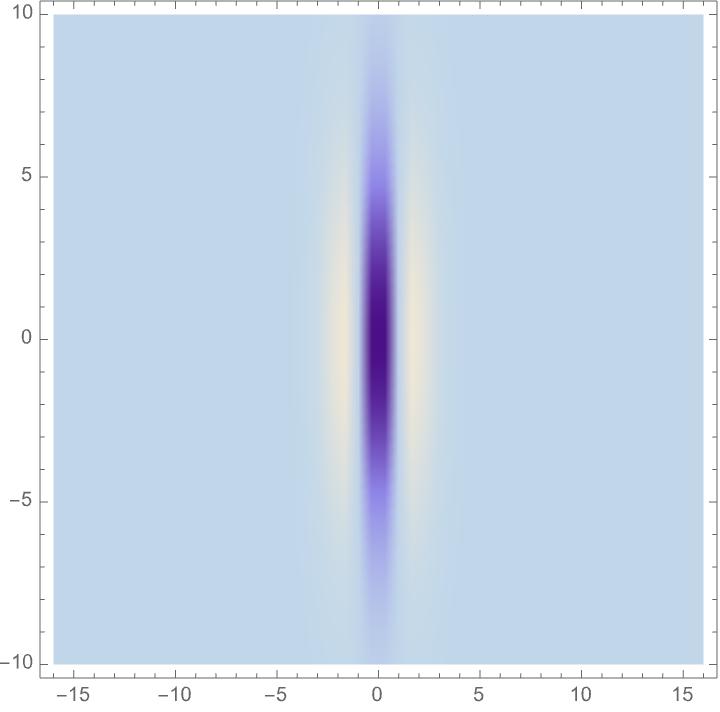}
      &  \includegraphics[width=0.14\textwidth]{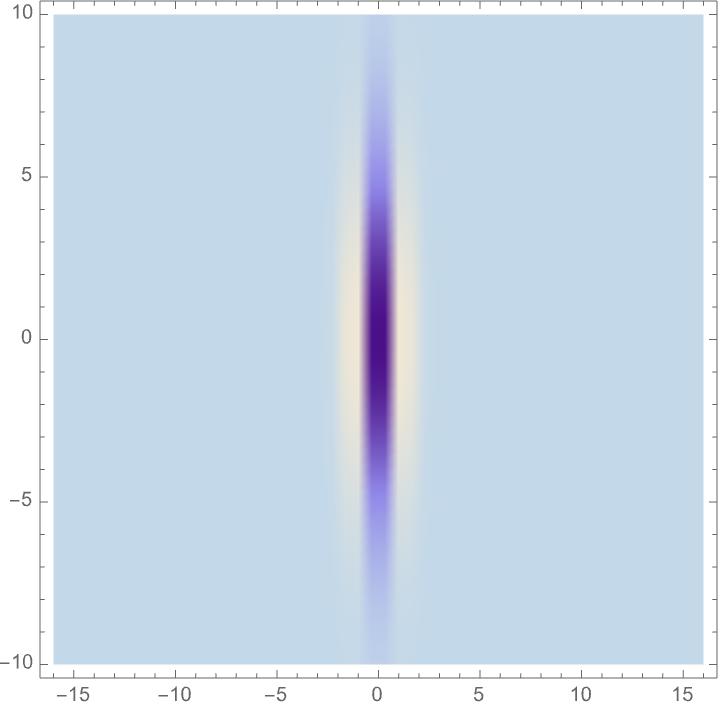} \\
       \includegraphics[width=0.14\textwidth]{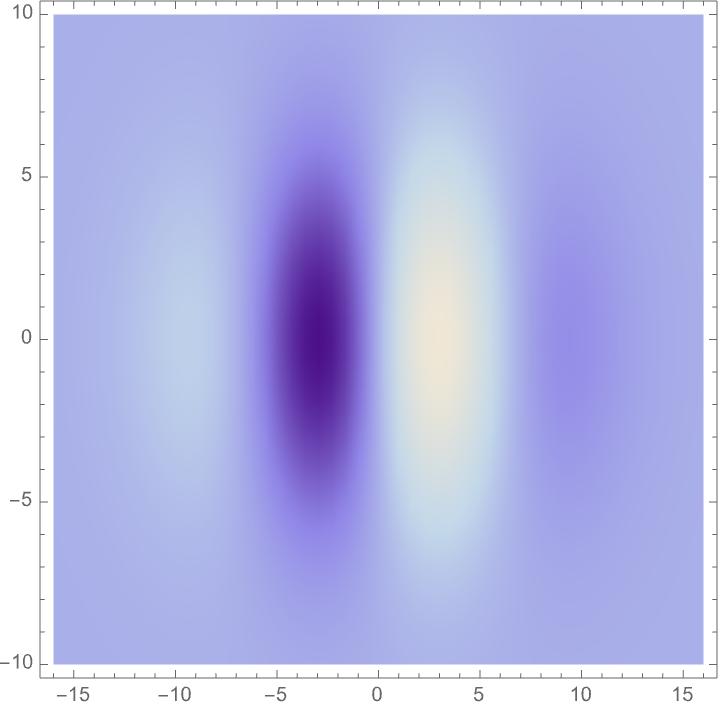}
      & \includegraphics[width=0.14\textwidth]{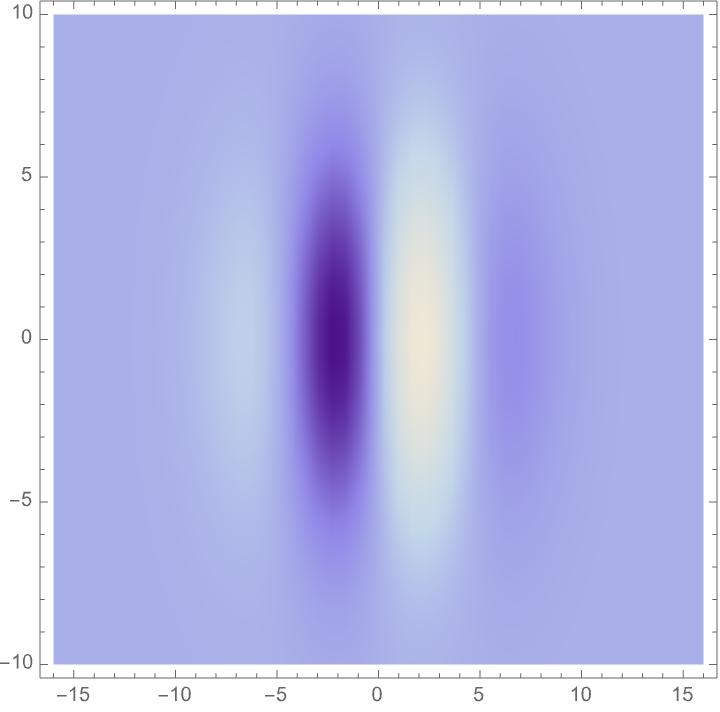}
      & \includegraphics[width=0.14\textwidth]{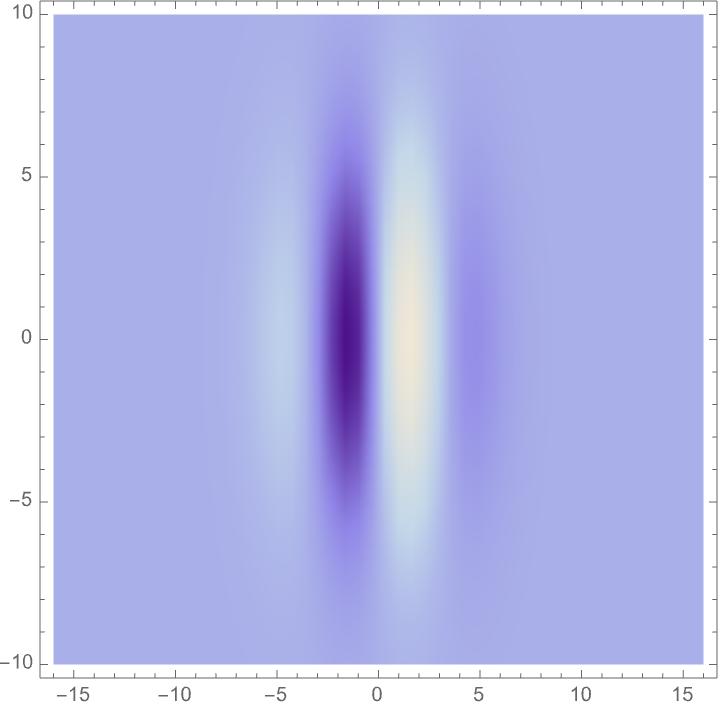}
      & \includegraphics[width=0.14\textwidth]{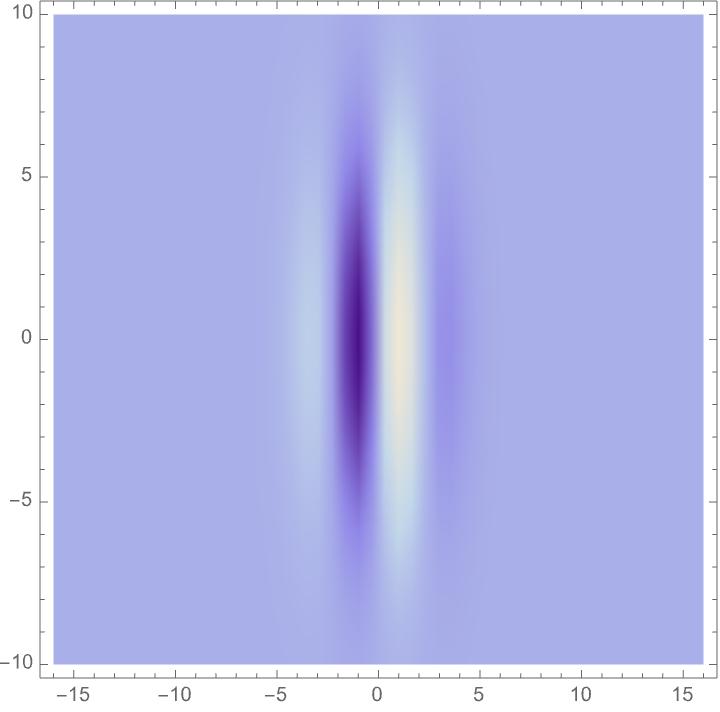}
      & \includegraphics[width=0.14\textwidth]{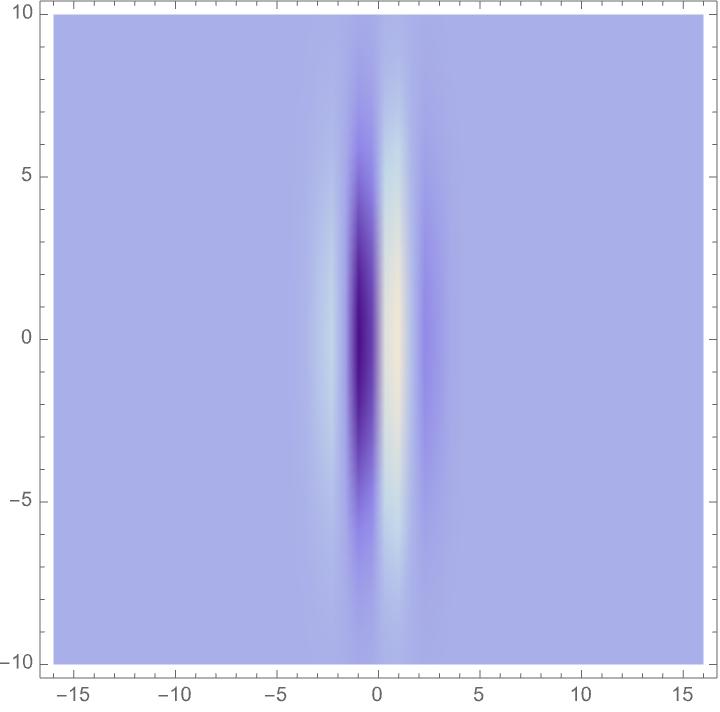}
      & \includegraphics[width=0.14\textwidth]{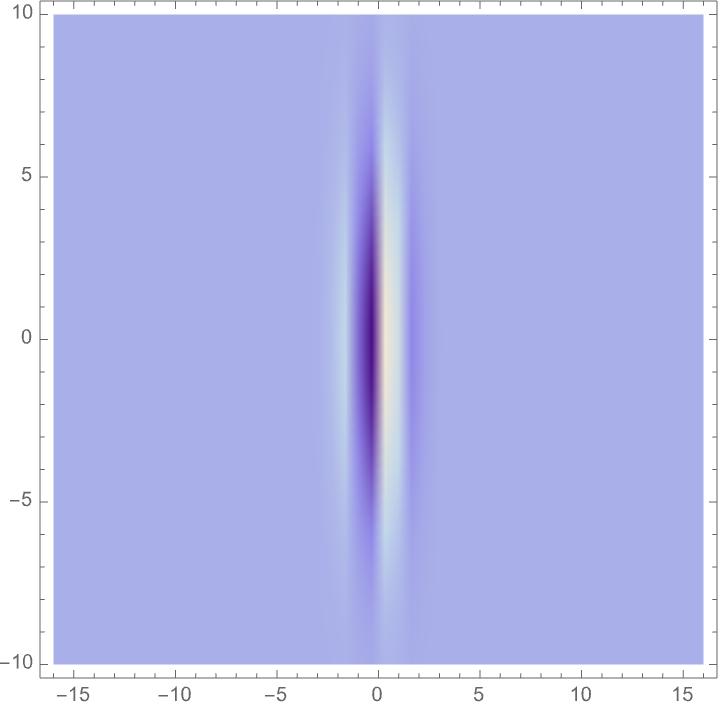} \\
      \includegraphics[width=0.14\textwidth]{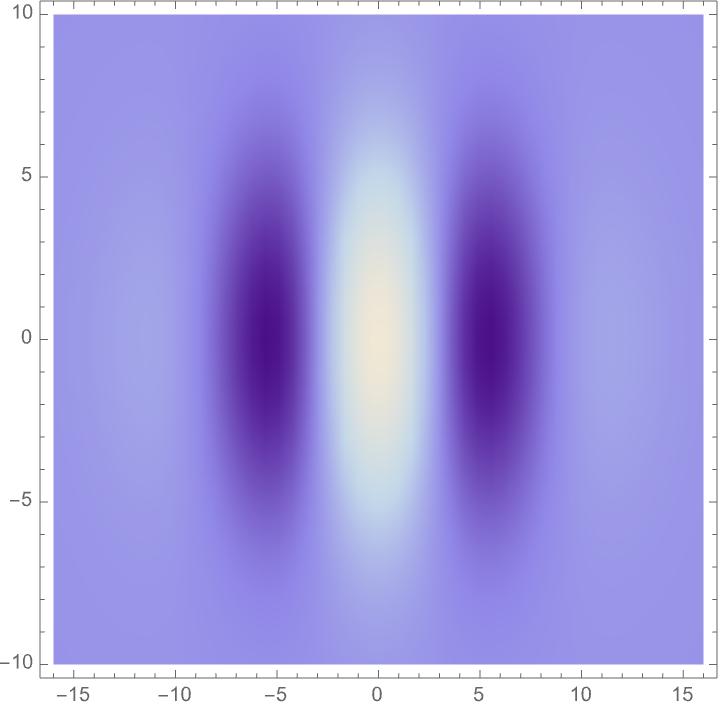}
      & \includegraphics[width=0.14\textwidth]{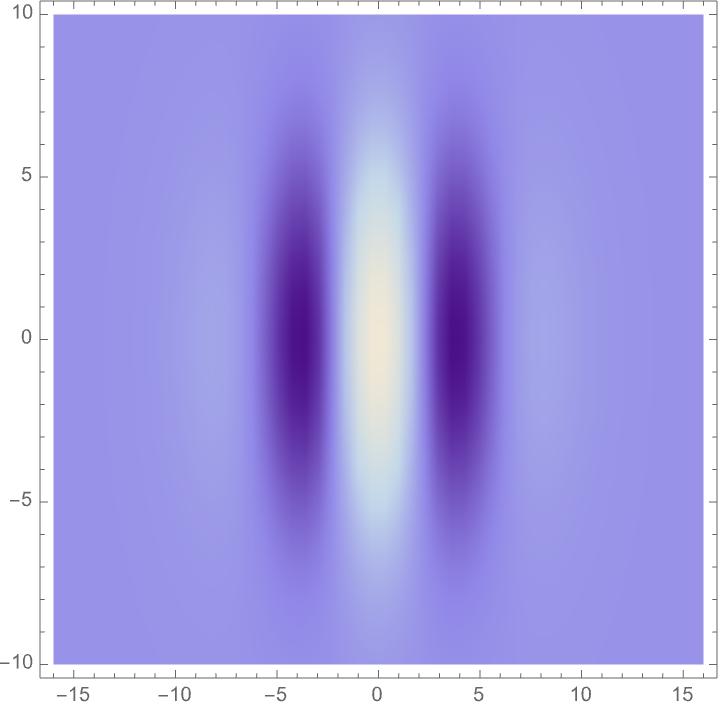}
      & \includegraphics[width=0.14\textwidth]{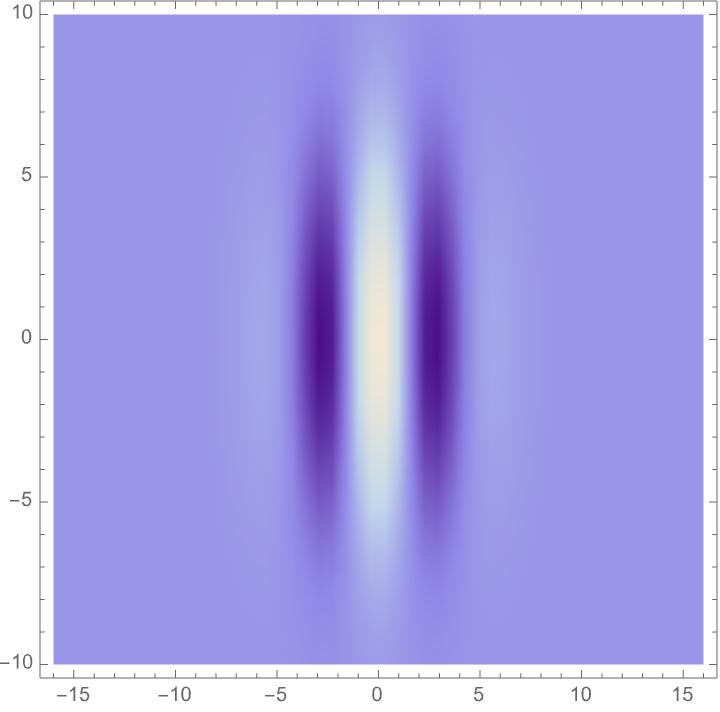}
      & \includegraphics[width=0.14\textwidth]{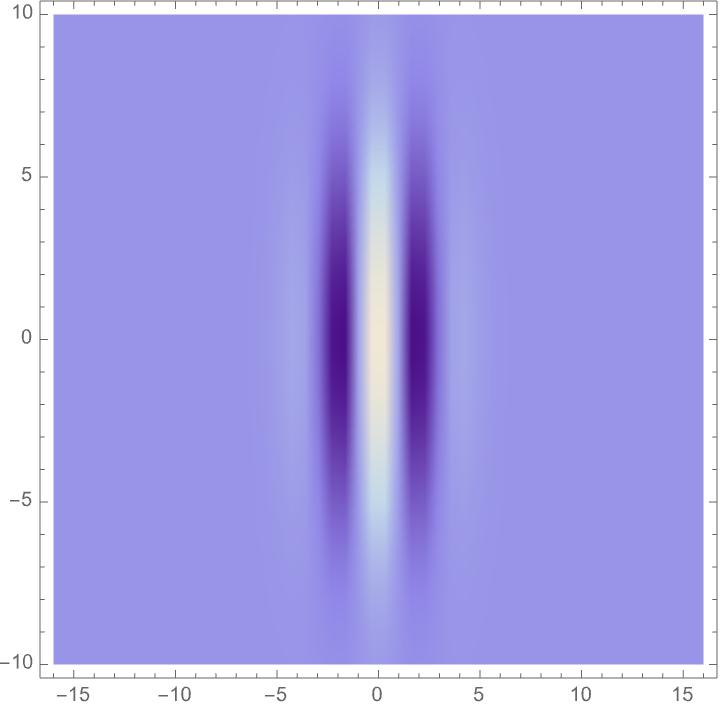}
      & \includegraphics[width=0.14\textwidth]{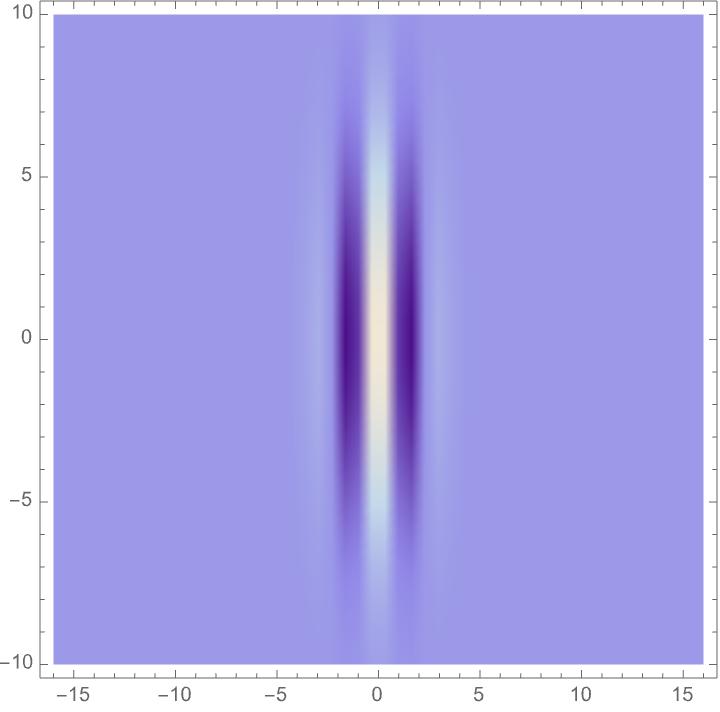}
      & \includegraphics[width=0.14\textwidth]{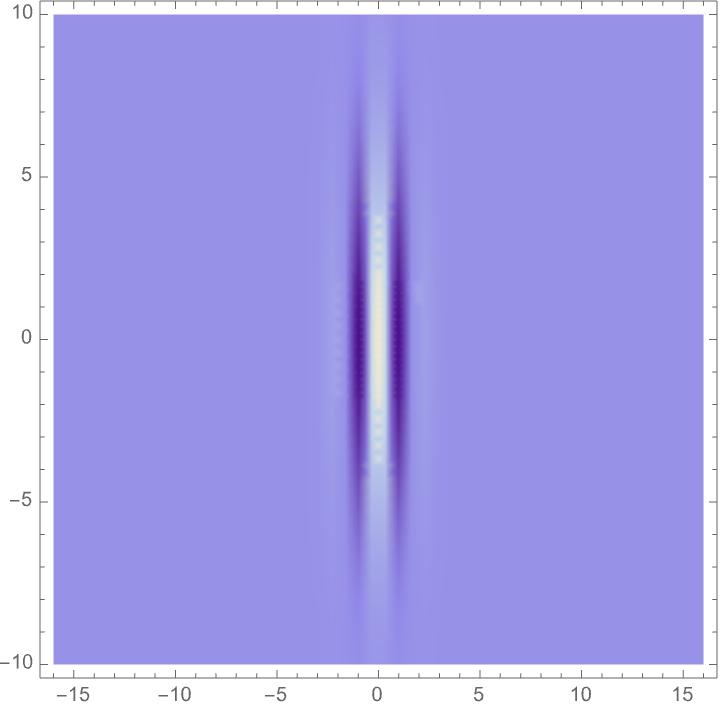}   
    \end{tabular}
  \end{center}
  \caption{Variability in the elongation of affine Gaussian derivative
    receptive fields (for the image orientation $\varphi = 0$),
    with the scale parameter ratio 
    $\kappa = \sigma_2/\sigma_1$ increasing from $1$ to $4\sqrt{2}$
    according to a logarithmic distribution, from left to right,
    with the vertical scale parameter kept constant $\sigma_2 = 4$
    and with the horizontal scale parameter being the smaller
    $\sigma_1 \leq \sigma_2$.
    (first row) First-order directional derivatives of affine Gaussian
    kernels according to (\ref{eq-spat-RF-model}) for $m = 1$.
    (second row) Second-order directional derivatives of affine
    Gaussian kernels according to (\ref{eq-spat-RF-model}) for $m = 2$.
    (third row) Second-order directional derivatives of affine
    Gaussian kernels according to (\ref{eq-spat-RF-model}) for $m = 3$.
    (fourth row) Second-order directional derivatives of affine
    Gaussian kernels according to (\ref{eq-spat-RF-model}) for $m = 4$.
    (Horizontal axes: image coordinate $x_1 \in [-16, 16]$.
     Vertical axes: image coordinate $x_2 \in [-16, 16]$.)}
  \label{fig-ecc-variability}
\end{figure*}

\section{Introduction}
\label{sec-intro}

When observing objects and events in our natural environment, the
image structures in the visual stimuli will be subject to substantial
variabilities caused by the natural image
transformations. Specifically, if observing a smooth local surface
patch from different viewing directions and viewing distances, this
variability can, to first order of approximation, be approximated by
local affine transformations (the derivative of the projective
mappings between the different views).
Within the degrees of freedom of 2-D spatial affine transformations%
\footnote{For an in-depth theoretical analysis of the degrees of
  freedom of 2-D spatial affine transformations in relation
  to the degrees of freedom of the shapes of 2-D visual
  receptive fields in terms of affine Gaussian derivatives, see
  Lindeberg (\citeyear{Lin24-arXiv-RelsDOFsaffGausstransf}).}
between two views of the same local surface patch, there
is one degree of freedom that corresponds to non-uniform
spatial scaling transformations with different amounts of
scaling along two orthogonal directions.
Geometrically, this degree of freedom corresponds
to variabilities in the slant angle between the local surface normal
and the viewing direction, as illustrated in
Figure~\ref{fig-castle-view-dirs} with multiple 
views of the same scene, obtained by moving the viewing
point and the viewing direction in a horizontal plane.

If we consider a vision system that views 3-D scenes from
multiple viewing directions, then the receptive field responses will
be strongly influenced by geometric image transformations.
A particular requirement on a vision system, as imposed
by the requirement of affine covariance
(to be described in more detail in
Sections~\ref{sec-aff-cov-rfs}--\ref{sec-hyp-aff-cov-rfs}),
is that the vision system should be able to match the outputs from the
receptive fields between different views of the same scene.
Then, under the foreshortening transformations that arise from
varying the slant angles under variations in the viewing conditions
of 3-D objects, the spatial shapes of the
receptive fields would have to be adapted to the resulting
non-uniform scaling transformations, to result in different
eccentricities of the receptive fields, as illustrated in 
Figure~\ref{fig-ecc-variability}.

In Lindeberg
(\citeyear{Lin21-Heliyon,Lin23-FrontCompNeuroSci,Lin24-arXiv-UnifiedJointCovProps,Lin24-arXiv-RelsDOFsaffGausstransf}),
we have outlined a general framework for how covariance properties with respect to
geometric image transformations may constitute a fundamental
constraint for the receptive fields in the primary visual cortex of
higher mammals. The underlying aim is to enable the visual
computations to be robust under
the variabilities in the image structures generated by the natural
image transformations. The fundaments of this theory are based on
axiomatically determined receptive field shapes, derived from symmetry
properties, that reflect structural properties of the environment, in
combination with additional constraints to guarantee consistency
between image representations over multiple spatial and temporal
scales.

The population of receptive fields in the primary visual
cortex ought to, according to this theory, obey covariance properties
with respect to spatial affine transformations and Galilean
transformations. Specifically, according to this theory,
the receptive field shapes ought to be expanded
over the degrees of freedom 
that correspond to the variabilities in image structures
caused by these classes of geometric image transformations.
Alternatively, the vision system ought to
implement computational mechanisms that would to
a sufficient good degree of approximation be computationally
equivalent to such an expansion of receptive field shapes
over the degrees of freedom of geometric image transformations.
Notably, the expansion of the number of receptive fields, from
about 1~M output channels from the lateral geniculate
nucleus to the primary visual cortex (V1) to 190~M
neurons in V1 with 37~M output channels
(see DiCarlo {\em et al.\/} (\citeyear{DiCZocRus12-Neuron}) Figure~3),
would indeed be consistent with an expansion of the receptive field
shapes in V1 over parameters of the receptive fields.

While overall qualitative comparisons between predictions from this principled theory have been successfully made to neurophysiological recordings of receptive fields of simple cells by DeAngelis {\em et al.\/}\
(\citeyear{DeAngOhzFre95-TINS,deAngAnz04-VisNeuroSci}),
Conway and Livingstone (\citeyear{ConLiv06-JNeurSci}) and
Johnson {\em et al.\/}\ (\citeyear{JohHawSha08-JNeuroSci}),
publicly available data regarding full receptive field recordings are
quite limited. Thus, further experimental evidence would be needed to
firmly either reject or support the stated hypotheses
about affine covariance and Galilean covariance.

In the lack of such neurophysiological data regarding full receptive
field recordings, one could, however, aim to instead obtain indirect
cues regarding a possible variability in the degree of elongation of
the receptive fields, by making use of the recordings of the
orientation selectivity of visual neurons in previous studies:
Experimental results by Nauhaus {\em et al.\/}\ (\citeyear{NauBenCarRin09-Neuron})
show a substantial variability regarding broad {\em vs.\/}\
sharp tuning of the receptive fields in the primary visual cortex.
Goris {\em et al.\/}\ (\citeyear{GorSimMov15-Neuron})
report comparably uniform distributions of the degree
of orientation selectivity of simple and complex cells,
in terms of histograms of the resultant of the orientation
selectivity curves.

In a companion paper (Lindeberg
\citeyear{Lin25-JCompNeurSci-orisel}), we have established a
strong direct link between the orientation selectivity and the
elongation of the receptive fields according to the idealized
generalized Gaussian derivative model for visual receptive fields (as
will be summarized in Section~\ref{sec-conn-ori-sel-elong}).
If the generalized Gaussian derivative shape is indeed a good model
for simple and complex cells in V1, then variations in their elongation
would logically correspond to a variations in their orientation
selectivity.

Thereby,
these results together are fully consistent with the hypothesis that the
visual receptive fields in the primary visual cortex should span%
\footnote{The terminology used throughout this paper of ``spanning a variability'',
  is similar to the terminology in mathematics concerning a 
  set of vectors that together span a vector space or a subspace.
  More generally, from the viewpoint of the wider classes of geometric
  image transformations in the generalized Gaussian derivative theory
  for visual receptive fields, one could consider these geometric
  image transformations to have different degrees of freedom.
  These degrees of freedom in the geometric image transformations
  will then correspond to different degrees of freedom in the
  shapes of the visual receptive fields, as induced by covariance
  properties of the receptive fields. A set of such degrees of freedom
  taken together will thereby span a corresponding variability in the
  shapes of the visual receptive fields.}
a variability in their anisotropy, thus consistent with the hypothesis
that the receptive fields should span at
least one more degree of freedom in the affine group, beyond mere
rotations (as can be established by the expansion of orientation
selectivity properties around the centers of the pinwheel structures
in the primary visual cortex of
higher mammals).

To ultimately judge is this hypothesis would hold, we will finally use
predictions from the presented theoretical analysis to formulate
a set of explicit, testable hypotheses, that
could be either verified or rejected in future neurophysiological experiments.
Additionally, we will formulate a set of quantitive measurements to be made,
to characterize a possible variability
in the anisotropy or elongation of receptive fields in the primary
visual cortex, with special emphasis on the relationships between a possibly
predicted variability in receptive field elongation and the pinwheel
structure in the primary visual cortex of higher mammals.

\subsection{Contributions and novelty}

In summary, the purposes of this paper are twofold:
\begin{itemize}
\item
  In the current absence of firm biological measurements about a
  possible variability of the degree of elongation of receptive field
  shapes in the primary visual cortex, provide potential indirect
  support for such a hypothesis. This reasoning is based
  on a combination of previously established variability in the degree
  of the orientation selectivity of biological receptive fields with a
  model-based connection between the degree of elongation of the
  receptive fields and their orientation selectivity.
\item
  To formulate a set of theoretically motivated and experimentally
  testable predictions and quantitative measurements, that could be
  used by experimentalists for ultimately judging whether the
  formulated hypothesis about a variability over the degree of
  elongation of the receptive fields would hold in the primary visual
  cortex of higher mammals. Such a connection could then provide
  potential further support for the more general hypothesis, that
  the receptive fields in the primary visual cortex of higher mammals may be covariant
  under the larger group of spatial affine transformations.
\end{itemize}
The main novel contributions of this paper are specifically:
\begin{itemize}
\item
  A qualitative explanation of the experimental results by
  Nauhaus {\em et al.\/}\ (\citeyear{NauBenCarRin09-Neuron})
  as the results of a systematic variability in the degree of
  elongation of the receptive fields.
\item
  Biological implications of that result, in that the
  pinwheel structure of higher mammals should,
  beyond an expansion over image orientations, could also comprise
  an expansion over the degree of elongation of the
  receptive fields, from the center of each pinwheel to its periphery.
\item
  Theoretical connections to, as well as potential partial support for,
  the hypothesis that the receptive fields in the primary visual
  cortex may be covariant under spatial affine transformations.
\item
  The quantitative explanation of the receptive field histograms
  obtained experimentally by Goris {\em et al.\/}\
  (\citeyear{GorSimMov15-Neuron}), as the results of
  combining the receptive fields of simple cells corresponding to
  different orders of spatial differentiation.
\item
  Biological implications of that result, that the experimentally
  recorded receptive field histograms are better explained by
  receptive fields in terms of spatial derivatives up to order 4 than
  in terms of spatial derivatives up to order 2.
\item
  The overall theoretical contributions in the paper, of explaining
  properties of the primary visual cortex, based on properties of a
  theoretically principled model for the receptive fields, using
  mathematical analysis and not numerical simulations of neural models
  as the main tool.
\end{itemize}
A further underlying motivation of this work is to lay out
a conceptual foundation, by which theoreticians and experimentalists could
join efforts to establish to what extent the distributions of the shapes
of the biological receptive fields would be compatible with an explanation from the
fundamental constraint, that the family of receptive fields
should be able to handle variabilities in
the image data caused by geometric image transformations.

\section{Related work}
\label{sec-rel-work}

Beyond the works by Nauhaus {\em et al.\/}\
(\citeyear{NauBenCarRin09-Neuron}) and by
Goris {\em et al.\/}\ (\citeyear{GorSimMov15-Neuron}),
that the treatment in Section~\ref{sec-results}
will largely build upon, there is a large body of work on
characterizing the orientation selectivity of neurons, by
Watkins and Berkley (\citeyear{WatBer73-ExpBrainRes}),  
Rose and Blakemore (\citeyear{RosBla74-ExpBrainRes}),
Schiller {\em et al.\/} (\citeyear{SchFinVol76-JNeuroPhys}),
Albright (\citeyear{Alb84-JNeuroPhys}),
Ringach {\em et al.\/} (\citeyear{RinShaHaw03-JNeurSci}),
Nauhaus {\em et al.\/}\ (\citeyear{NauBenCarRin09-Neuron}),
Scholl {\em et al.\/} (\citeyear{SchTanCorPri13-JNeurSci}),
Sadeh and Rotter (\citeyear{SadRot14-BICY})
and
Sasaki {\em et al.\/}
(\citeyear{SakKimNimTabTanFukAsaAraInaNakBabDaiNisSanTanImaTanOhz15-SciRep}),
as well as concerning biological mechanisms for achieving orientation selectivity by
Somers {\em et al.\/}\ (\citeyear{SomNelSur95-JNeuroSci}),
Sompolinsky and Shapley (\citeyear{SomSha97-CurrOpNeuroBio}),
Carandini and Ringach (\citeyear{CarRin97-VisRes}),
Lampl {\em et al.\/}\ (\citeyear{LamAndGilFer01-Neuron}),
Ferster and Miller (\citeyear{FerMil00-AnnRevNeuroSci}),
Shapley {\em et al.\/}\ (\citeyear{ShaHawRin03-Neuron}),
Seri{\`e}s {\em et al.\/}\ (\citeyear{SerLatPou04-NatNeuroSci}),
Hansel and van~Vreeswijk (\citeyear{HanVre12-JNeuroSci}),
Moldakarimov {\em et al.\/}\ (\citeyear{MolBazSej14-PLOSCompBiol}),
Gonzalo Cogno and Mato (\citeyear{GonMat15-FrontNeurCirc}),
Priebe (\citeyear{Pri16-AnnRevVisSci}),
Pattadkal {\em et al.\/} (\citeyear{PatMatVrePriHan18-CellRep}),
Nguyen and Freeman (\citeyear{NguFre19-PLOSCompBiol}),
Merkt {\em et al.\/} (\citeyear{MerSchRot19-PLOSCompBiol}),
Wei {\em et al.\/} (\citeyear{WeiMerRot22-bioRxiv})
and
Wang {\em et al.\/} (\citeyear{WanDeyLagBehCalSta24-CellRep}).
The focus of this paper, however, is not on the neural mechanisms that lead to orientation selectivity, but on purely {\em functional properties\/} at the macroscopic level.

Mathematical models of biological receptive fields have been
formulated in terms of Gaussian derivatives
(Koenderink and van Doorn \citeyear{Koe84,KoeDoo87-BC,KoeDoo92-PAMI};
 Young and his co-workers \citeyear{You87-SV,YouLesMey01-SV,YouLes01-SV};
 Lindeberg \citeyear{Lin13-BICY,Lin21-Heliyon})
and Gabor filters
(Marcelja \citeyear{Mar80-JOSA};
Jones and Palmer \citeyear{JonPal87a,JonPal87b};
Porat and Zeevi \citeyear{PorZee88-PAMI}).
Gaussian derivatives have, in turn, been used as primitives in theoretical models of early visual processing by
Lowe (\citeyear{Low00-BIO}),
May and Georgeson (\citeyear{MayGeo05-VisRes}),
Hesse and Georgeson (\citeyear{HesGeo05-VisRes}),
Georgeson  {\em et al.\/}\ (\citeyear{GeoMayFreHes07-JVis}),
Hansen and Neumann (\citeyear{HanNeu09-JVis}),
Wallis and Georgeson (\citeyear{WalGeo09-VisRes}),  
Wang and Spratling (\citeyear{WanSpra16-CognComp}),
Pei {\em et al.\/}\ (\citeyear{PeiGaoHaoQiaAi16-NeurRegen}),
Ghodrati {\em et al.\/}\ (\citeyear{GhoKhaLeh17-ProNeurobiol}),
Kristensen and Sandberg (\citeyear{KriSan21-SciRep}),
Abballe and Asari (\citeyear{AbbAsa22-PONE}),
Ruslim {\em et al.\/}\ (\citeyear{RusBurLia23-bioRxiv}) and
Wendt and Faul (\citeyear{WenFay24-JVis}).

Hubel and Wiesel
(\citeyear{HubWie59-Phys,HubWie62-Phys,HubWie68-JPhys,HubWie05-book})
pioneered the study of simple and complex cells.
The properties of simple cells have been further characterized by
DeAngelis {\em et al.\/}\
(\citeyear{DeAngOhzFre95-TINS,deAngAnz04-VisNeuroSci}),
Ringach (\citeyear{Rin01-JNeuroPhys,Rin04-JPhys}),
Conway and Livingstone (\citeyear{ConLiv06-JNeurSci}),
Johnson {\em et al.\/}\ (\citeyear{JohHawSha08-JNeuroSci}),
Walker {\em et al.\/}
(\citeyear{WalSinCobMuhFroFahEckReiPitTol19-NatNeurSci})
and
De and Horwitz (\citeyear{DeHor21-JNPhys}).
Properties of complex cells have been investigated by
Movshon {\em et al.\/}\ (\citeyear{MovThoTol78-JPhys}), 
Emerson {\em et al.\/}\ (\citeyear{EmeCitVauKle87-JNeuroPhys}),
Martinez and Alonso (\citeyear{MarAlo01-Neuron}),
Touryan {\em et al.\/}\ (\citeyear{TouLauDan02-JNeuroSci,TouFelDan05-Neuron}),
Rust {\em et al.\/}\ (\citeyear{RusSchMovSim05-Neuron}),
van~Kleef {\em et al.\/}\ (\citeyear{KleCloIbb10-JPhys}),  
Goris {\em et al.\/}\ (\citeyear{GorSimMov15-Neuron}),
Li {\em et al.\/}\ (\citeyear{LiLiuChoZhaTao15-JNeuroSci}) and
Almasi {\em et al.\/}\ (\citeyear{AlmMefCloWonYunIbb20-CerCort}),
as well as modelled computationally by
Adelson and Bergen (\citeyear{AdeBer85-JOSA}),
Heeger (\citeyear{Hee92-VisNeuroSci}),
Serre and Riesenhuber (\citeyear{SerRie04-AIMemo}),
Einh{\"a}user {\em et  al.\/} (\citeyear{EinKayKoeKoe02-EurJNeurSci}),
Kording {\em et al.\/}\ (\citeyear{KorKayWinKon04-JNeuroPhys}),
Merolla and Boahen (\citeyear{MerBoa04-NIPS}),
Berkes and Wiscott (\citeyear{BerWis05-JVis}),
Carandini (\citeyear{Car06-JPhys}),
Hansard and Horaud (\citeyear{HanHor11-NeurComp}),
Franciosini {\em et al.\/}\ (\citeyear{FraBouPer19-AnnCompNeurSciMeet}),
Lindeberg (\citeyear{Lin20-JMIV}),
Lian {\em et  al.\/} (\citeyear{LiaAlmGraKamBurMef21-PLOSCompBiol}),
Oleskiw {\em et al.\/}\ (\citeyear{OleLieSimMov23-bioRxiv})
and Yedjour and Yedjour (\citeyear{YedYed24-CognNeurDyn}).
In this work, we will follow a specific way of modelling simple and complex cells in terms of affine Gaussian derivatives, according to the generalized affine Gaussian derivative model for visual receptive fields.

Properties of cortical maps in the primary visual cortex have, in
turn, been studied in detail by
Bonhoeffer and Grinvald (\citeyear{BonGri91-Nature}),
Blasdel (\citeyear{Bla92-JNeuroSci}),
Maldonado {\em et al.\/} \citeyear{MalGodGraBon97-Science},
Koch {\em et al.\/} (\citeyear{KocJinAloZai16-NatComm}),
Kremkow {\em et al.\/}\ (\citeyear{KreJinWanAlo16-Nature}),
Najafian {\em et al.\/}\
(\citeyear{NajKocTehJinRahZaiKreAlo22-NatureComm}),
Jung {\em et al.\/}\ (\citeyear{JunAlmSunYunCloBauRenMefIbb22-SciAdv}),
Fang {\em et al.\/}\ (\citeyear{FanCaiLu-PNAS}) and
Vita  {\em et al.\/}\ (\citeyear{BitOrsStaClaTir24-bioRxiv}).
 
There have been previous mentions regarding receptive fields with
different aspect ratios 
(Tinsley {\em et al.\/}\ \citeyear{TinWebBarVinParDer03-JNeuroPhys},
Xu  {\em et al.\/}\ \citeyear{XuLiCheWanYan16-CognNeurDyn}).
It has also been reported by Wilson {\em et al.\/}
(\citeyear{WilWhiSchFit16-NatNeuroSci})
that ``neurons located near pinwheel centers
... exhibit broader orientation tuning than neurons in regions of the
map where neighbouring neurons exhibit similar preferences''.

According to our knowledge, there have, however, not been any
previous in-depth studies in
relation to a systematic expansions of receptive field shapes
over the degree of elongation, nor any
previously reported connections between such expansions of
receptive field shapes over the degree of elongation
in relation to the property of affine
covariance for the family of visual receptive fields.

\section{Theoretical background}

In this section, we: (i)~give a theoretical background regarding the
notion of affine covariant visual receptive fields, which constitutes
the conceptual background for the hypotheses studied in this work;
(ii)~describe how the orientation selectivity of receptive fields is
related to the degree of elongation of the receptive fields, based on
an in-depth theoretical analysis of visual receptive fields according
to the generalized Gaussian derivative model; and
(iii)~relate the computational modelling approaches taken and the contributions presented
in this work to previous work in the field.

The material in this section will then constitute the conceptual
background for the novel contributions in the paper, to be presented
in Section~\ref{sec-results}, such that the
paper can be read in a self-contained manner, without necessarily
having to first read the references Lindeberg
(\citeyear{Lin23-FrontCompNeuroSci}, \citeyear{Lin25-JCompNeurSci-orisel}),
which this paper closely
builds upon.


\subsection{Affine covariant visual receptive fields}
\label{sec-aff-cov-rfs}

Let us represent the spatial image coordinates by $x = (x_1, x_2)^T \\
\in \bbbr^2$ and centered
affine spatial transformations in the 2-D image domain as
\begin{equation}
  x' = A \, x,
\end{equation}
where $A$ represents any non-singular $2 \times 2$ matrix
and $x' = (x'_1, x'_2)^T \in \bbbr^2$ denotes the transformed image coordinates.

Then, an affine transformed image
$f' \colon \bbbr^2 \rightarrow \bbbr$ of an original image
$f' \colon \bbbr^2 \rightarrow \bbbr$ is defined according to
\begin{equation}
  f'(x') = f(x).
\end{equation}
Let us denote the space of sufficiently smooth functions
corresponding to continuous image data $\bbbr^2 \rightarrow \bbbr$ by $V$.
With the affine transformation operator ${\cal T}_A : V \rightarrow V$
from this space onto itself, we can write the affine transformation
as
\begin{equation}
   f' = {\cal T}_A \, f.
\end{equation}
The property of affine covariance then means that the results of either:
\begin{itemize}
\item
  applying an affine transformation $x' = A \, x$ to an image
  $f$ and then applying a receptive field ${\cal R}' : V \rightarrow V$ to the affine
  transformed image $f'$, or
\item
  applying a related receptive field ${\cal R} : V \rightarrow V$ to the original image
  $f$ and then applying an affine transformation ${\cal T}_A$ to that output,
\end{itemize}
will lead to the same result, such that
\begin{equation}
   {\cal R}' \, {\cal T}_A\, f = {\cal T}_A \, {\cal R} \, f,
\end{equation}
where the affine covariant property of the receptive field family means
that for every receptive field ${\cal R}$ in the receptive field
family, there exists a possibly transformed receptive field
${\cal R}'$ within the same family, specifically determined
according to the actual value of the affine transformation matrix
$A$, such that the above relationship is guaranteed to hold, for some
transformed receptive field ${\cal R}'$ as a function of the original
receptive field ${\cal R}$ and the affine transformation matrix $A$.

The property of affine covariance thus means that the family of
receptive fields is well-behaved with regard to spatial affine
transformations, in the sense that affine transformations commute with
the operation of computing outputs from the family of receptive
fields.

In  Lindeberg (\citeyear{Lin21-Heliyon,Lin23-FrontCompNeuroSci,Lin24-arXiv-UnifiedJointCovProps}), it
is argued that such affine covariant properties constitute an
essential property of spatial receptive fields, as well as for the spatial
components in joint spatio-temporal receptive fields. Specifically,
the receptive fields, according to the generalized Gaussian derivative
model for visual receptive fields, to be used below,
obey such affine covariant properties.

The property of the degree of elongation of the receptive
fields, to be studied in detail in this work, 
spans a 1-D variability within the full 4-D variability of general
affine transformations. Thus, it constitutes one of the degrees of freedom in
the variability of transformed receptive field shapes of ${\cal R}'$, that will be
generated by subjecting an original receptive field ${\cal R}$ to the
4-D variability of general affine transformation matrices $A$;
see Lindeberg (\citeyear{Lin24-arXiv-RelsDOFsaffGausstransf}) for further details.

\subsection{The hypothesis about affine covariant receptive fields}
\label{sec-hyp-aff-cov-rfs}

If we assume that the visual system should implement affine covariant
receptive fields (Lindeberg \citeyear{Lin23-FrontCompNeuroSci} Section~3.2),
then the property of affine covariance would make
it possible to compute better estimates of local surface orientation,
compared to a visual system that does not implement affine covariance,
or a sufficiently good approximation thereof.

A general motivation for the wider underlying working hypothesis
about affine covariance can be stated as follows: If the
population of receptive fields would support affine covariance in the
primary visual cortex, or sufficiently good approximations thereof,
then such an ability would support the possibility of computing
affine invariant image representations at higher levels in the visual
hierarchy (Lindeberg \citeyear{Lin13-PONE}).
Alternatively, one could consider also sufficiently good approximations
thereof, over restricted subspaces or subdomains of the most general
forms of full variability 
under spatial affine transformations of the
visual stimuli.

Fundamentally, we cannot expect the visual perception system to implement
full affine invariance.
For example, from the well-known experience, that it is much harder to read text
upside-down, it is clear that the visual perception system cannot be
regarded as
  invariant to spatial rotations in the image domain. However, from the
  expansion of the orientations of visual receptive fields according
  to the pinwheel structure of higher mammals, we can regard
  the population of receptive fields as supporting local rotational
  covariance.
  
When we look
  at a slanted surface in the world, we can get a robust and stable
  perception of its surface texture under substantial variations of
  the slant angle. This robustness of visual perception 
  under stretchings of image patterns that correspond to
  non-uniform scaling transformations
  (the perspective effects on
  a slanted surface patch can, to first-order of approximation, be modelled as
  a stretching of the image pattern along the tilt direction in image
  space, complemented with a uniform scaling transformation).
  If the visual receptive fields would span a variability under such spatial
  stretching transformations, then such a variability would precisely
  correspond to a variability in the anisotropy, or the degree of
  elongation, of the receptive fields.

From these theoretical motivations, one may hence raise the question
whether biological vision in higher mammals would have developed
computational mechanisms in the visual pathway that could be modelled
in terms of affine covariance. In Lindeberg
(\citeyear{Lin23-FrontCompNeuroSci}, Hypothesis~1 in Section~3.2.1),
it has been proposed whether the visual
receptive fields in the primary visual cortex may comprise an
``expansion of spatial receptive field shapes over a larger part of
the affine group than mere rotations or uniform scale changes''.

The main topic of this paper is to investigate the possible validity of
this hypothesis with respect to the degree of freedom of spatial affine
transformations corresponding to non-uniform scaling transformations
over the image domain, and corresponding to receptive field shapes of
different degrees of elongation.

\subsection{Connections between the orientation selectivity and the
  degree of elongation of the receptive fields for the generalized
  Gaussian derivative model for visual receptive fields}
\label{sec-conn-ori-sel-elong}

For modelling the receptive fields in the primary visual cortex, we
will use the generalized Gaussian derivative model for
receptive fields (Lindeberg \citeyear{Lin21-Heliyon}).

\subsubsection{Idealized models for simple cells}

We will model the purely spatial component of the receptive fields for simple cells as
(Lindeberg \citeyear{Lin21-Heliyon} Equation~(23); see Figure~7 in that reference for illustrations)
\begin{multline}
  \label{eq-spat-RF-model}
  T_{\simple}(x_1, x_2;\; \sigma_{\varphi}, \varphi, \Sigma_{\varphi}, m) \\
  = T_{\varphi^m,\norm}(x_1, x_2;\; \sigma_{\varphi}, \Sigma_{\varphi})
  = \sigma_{\varphi}^{m} \, \partial_{\varphi}^{m} \left( g(x_1, x_2;\; \Sigma_{\varphi}) \right),
\end{multline}
and with joint spatio-temporal receptive fields of the simple cells according to
(Lindeberg \citeyear{Lin21-Heliyon} Equation~(25); see Figures~10-11 in that reference for illustrations)
\begin{align}
  \begin{split}
    \label{eq-spat-temp-RF-model-der-norm-caus}
    T_{\simple}(x_1, x_2, t;\; \sigma_{\varphi}, \sigma_t, \varphi, v, \Sigma_{\varphi}, m, n) 
  \end{split}\nonumber\\
  \begin{split}
   & = T_{{\varphi}^m, {\bar t}^n,\norm}(x_1, x_2, t;\; \sigma_{\varphi}, \sigma_t, v, \Sigma_{\varphi})
  \end{split}\nonumber\\
  \begin{split}
   &  = \sigma_{\varphi}^{m} \, 
          \sigma_t^{n} \, 
         \partial_{\varphi}^{m} \,\partial_{\bar t}^n 
          \left( g(x_1 - v_1 t, x_2 - v_2 t;\; \Sigma_{\varphi}) \,
           h(t;\; \sigma_t) \right),
  \end{split}
\end{align}
where
\begin{itemize}
\item
   $\varphi \in [-\pi, \pi]$ is the preferred orientation of the receptive
   field,
\item
  $\sigma_{\varphi} \in \bbbr_+$ is the amount of spatial smoothing,
\item
  $\partial_{\varphi}^m =
  (\cos \varphi \, \partial_{x_1} + \sin  \varphi \, \partial_{x_2})^m$
  is an $m$:th-order directional derivative operator,
   in the direction $\varphi$,
 \item
   $\Sigma_{\varphi}$ is a $2 \times 2$ symmetric positive definite covariance matrix, with
   one of its eigenvectors in the direction of $\varphi$, 
 \item
   $g(x;\; \Sigma_{\varphi})$ is a 2-D affine Gaussian kernel with its shape
   determined by the covariance matrix $\Sigma_{\varphi}$
   \begin{equation}
     g(x;\; \Sigma_{\varphi})
     = \frac{1}{2 \pi \sqrt{\det \Sigma_{\varphi}}}
         e^{-x^T \Sigma_{\varphi}^{-1} x/2}
    \end{equation}
    for $x = (x_1, x_2)^T \in \bbbr^2$, 
\item
  $\sigma_t \in \bbbr_+$ is the amount of temporal smoothing,
\item
  $v = (v_1, v_2)^T \in \bbbr^2$ is a local motion vector, in the
  direction $\varphi$ of the spatial orientation of the receptive field,
\item
  $\partial_{\bar t}^n = (\partial_t + v_1 \, \partial_{x_1} + v_2 \, \partial_{x_2})^n$
  is an $n$:th-order velocity-adapted temporal derivative
  operator, and
\item
  $h(t;\; \sigma_t)$ is a temporal Gaussian kernel with 
  standard deviation $\sigma_t$ over time $t \in \bbbr$.
\end{itemize}
This model builds upon the regular Gaussian derivative model for purely spatial receptive fields proposed by Koenderink and van Doorn (\citeyear{Koe84,KoeDoo87-BC,KoeDoo92-PAMI})
and previously used for modelling biological fields by
Young and his co-workers (\citeyear{You87-SV,YouLesMey01-SV,YouLes01-SV}).
Here, that regular Gaussian derivative model is additionally generalized to
affine covariance, according to Lindeberg (\citeyear{Lin13-BICY,Lin21-Heliyon}).

\subsubsection{Idealized models for complex cells}

To model complex cells with a purely spatial dependency, we will use a quasi-quadrature measure of the form (Lindeberg \citeyear{Lin20-JMIV} Equation~(39))
\begin{equation}
  \label{eq-quasi-quad-dir}
  {\cal Q}_{\varphi,\spat,\norm} L
  = \sqrt{L_{\varphi,\norm}^2
               + C_{\varphi} \, L_{\varphi\varphi,\norm}^2},
\end{equation}
where
\begin{itemize}
\item
  $L_{\varphi,\norm}$ and $L_{\varphi\varphi,\norm}$ constitute the results of convolving the input image with scale-normalized directional affine Gaussian derivative operators of orders 1 and~2:
  \begin{multline}
       L_{\varphi,\norm}(x_1, x_2;\;  \sigma_{\varphi}, \Sigma_{\varphi}) =\\
       = T_{\varphi,\norm}(x_1, x_2;\; \sigma_{\varphi}, \Sigma_{\varphi}) *
       f(x_1, x_2),
    \end{multline}
    \begin{multline}
       L_{\varphi\varphi,\norm}(x_1, x_2;\;  \sigma_{\varphi}, \Sigma_{\varphi}) =\\
        = T_{\varphi\varphi,\norm}(x_1, x_2;\; \sigma_{\varphi}, \Sigma_{\varphi}) *
       f(x_1, x_2),
    \end{multline}
\item
  $C_{\varphi} > 0$ is a weighting factor between first and second-order
  information.
\end{itemize}
This model constitutes an affine Gaussian derivative analogue of
the energy model of complex cells developed by
Adelson and Bergen (\citeyear{AdeBer85-JOSA}) and
Heeger (\citeyear{Hee92-VisNeuroSci}), and is consistent with the observation that
receptive fields analogous to
first- {\em vs.\/}\ second-order derivatives occur in
pairs in biological vision (De~Valois {\em et al.\/}\ \citeyear{ValCotMahElfWil00-VR}),
with close analogies to quadrature pairs, as defined in terms of a
Hilbert transform  (Bracewell \citeyear{Bra99}, pp.~267--272).

Complex cells with a joint spatio-temporal dependency will, in turn, be modelled as
\begin{multline}
  \label{eq-quasi-quad-dir-vel-adapt-spat-temp}
  ({\cal Q}_{\varphi,\vel,\norm} L)
  = \sqrt{L_{\varphi,\norm}^2 
              + \, C_{\varphi} \, L_{\varphi\varphi,\norm}^2},
\end{multline}
where
\begin{multline}
  \label{eq-spat-temp-quasi-vel-adapt-comp1}
    L_{\varphi,\norm}(x_1, x_2, t;\;  \sigma_{\varphi}, \sigma_t, v, \Sigma_{\varphi}) =\\
    = T_{\varphi,\norm}(x_1, x_2, t;\; \sigma_{\varphi}, \sigma_t, v, \Sigma_{\varphi}) *
        f(x_1, x_2, t),
 \end{multline}
\begin{multline}
    \label{eq-spat-temp-quasi-vel-adapt-comp2}
    L_{\varphi\varphi,\norm}(x_1, x_2, t;\;  \sigma_{\varphi}, \sigma_t, v, \Sigma_{\varphi}) =\\
    = T_{\varphi\varphi,\norm}(x_1, x_2, t;\; \sigma_{\varphi}, \sigma_t, v, \Sigma_{\varphi}) *
        f(x_1, x_2, t),
\end{multline}
with the underlying space-time separable spatio-temporal receptive fields
$T_{\varphi^m, t^n,\norm}(x_1, x_2, t;\; \sigma_{\varphi}, \sigma_t, v, \Sigma_{\varphi}) $
according to (\ref{eq-spat-temp-RF-model-der-norm-caus}) for $n = 0$.

\begin{figure}[hbtp]
  \begin{center}
    \includegraphics[width=0.35\textwidth]{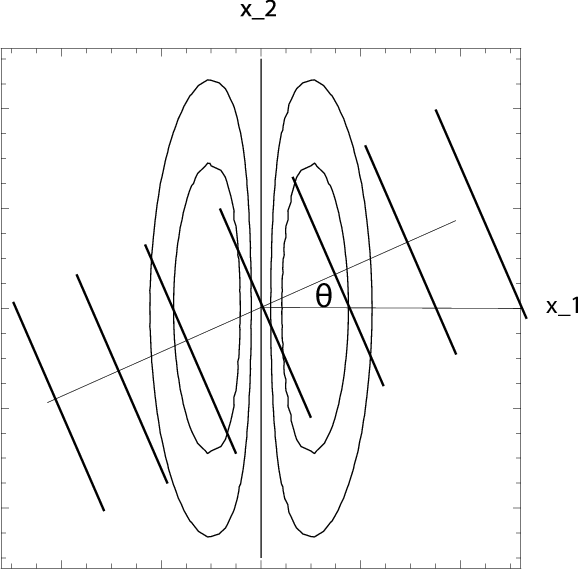}
  \end{center}
  \caption{Schematic illustration of the sine wave probe used for defining the orientation selectivity curve, by using a receptive field model with the fixed
    preferred orientation
    $\varphi = 0$, and then exposing the receptive field to sine waves
    for different inclination angles $\theta$.
    (Horizontal axis: spatial coordinate $x_1$.
  Vertical axis: spatial coordinate $x_2$.)}
  \label{fig-schem-ill-model-rf-sine}
\end{figure}

\begin{figure*}[hbtp]
  \begin{center}
       \begin{tabular}{ccc}
      {\em\footnotesize First-order simple cells\/}
      &       {\em\footnotesize Second-order simple cells\/}
      &       \hspace{-14mm} {\em\footnotesize Complex cells\/} \\
      \includegraphics[height=0.18\textwidth]{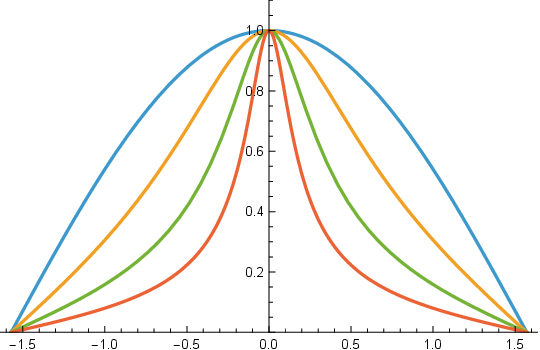}
      & \includegraphics[height=0.18\textwidth]{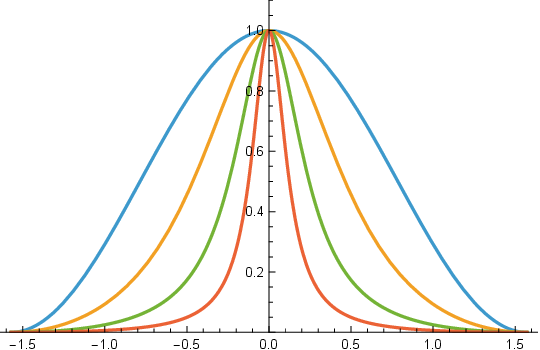}
      &
        \includegraphics[height=0.18\textwidth]{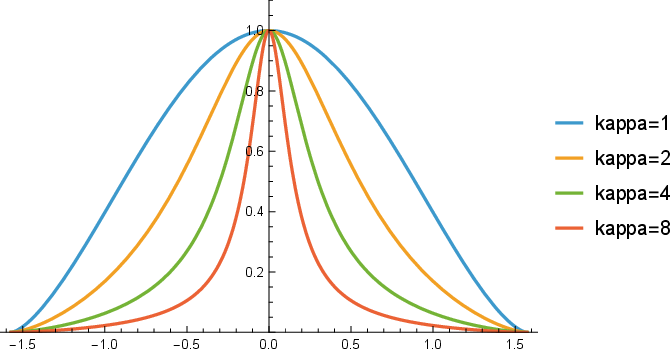}
         \\ $\,$ \\
        {\em\footnotesize Third-order simple cells\/}
      &   {\em\footnotesize Fourth-order simple cells\/} \\
      \includegraphics[height=0.18\textwidth]{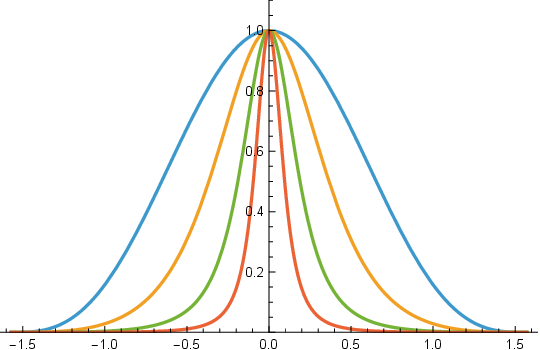}
      &
        \includegraphics[height=0.18\textwidth]{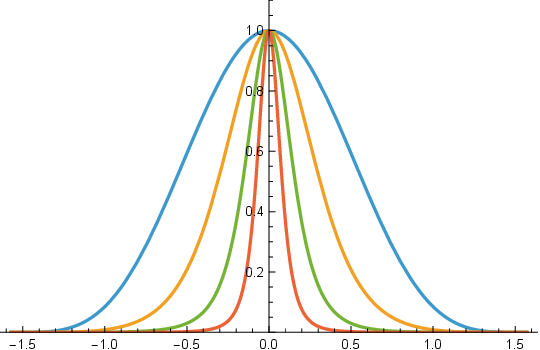} \\
      \end{tabular}
  \end{center}
  \caption{Graphs of the orientation selectivity for the idealized models of
   (top left) simple cells in terms of first-order
    directional derivatives of affine Gaussian kernels, (top middle) simple
    cells in terms of second-order directional derivatives of affine Gaussian
    kernels, (top right) complex cells in terms of directional
    quasi-quadrature measures that combine the first- and second-order
    simple cell responses in a Euclidean way for $C_{\varphi} = C_t =
    1/\sqrt{2}$,
    (bottom left) simple cells in terms of third-order
    directional derivatives of affine Gaussian kernels and (bottom middle) simple
    cells in terms of fourth-order directional derivatives of affine Gaussian
    kernels, and shown for different values of the ratio
    $\kappa$ between the spatial scale parameters in the vertical
    {\em vs.\/}\ the horizontal directions. Observe how the degree of orientation
    selectivity varies strongly depending on the eccentricity
    $\epsilon = 1/\kappa$ of the receptive fields.
    (Horizontal axes: orientation $\theta \in [-\pi/2, \pi/2]$.
     Vertical axes: Amplitude of the receptive field response relative
     to the maximum response obtained for $\theta = 0$.)}
  \label{fig-ori-sel}
\end{figure*}

\subsubsection{Orientation selectivity curves for the idealized receptive field models}
\label{sec-ori-sel-curves-ideal-rf-models}

In Lindeberg (\citeyear{Lin25-JCompNeurSci-orisel}),
the responses of the above purely spatial models of receptive fields
are calculated with respect to a static sine wave with orientation
$\theta \in [-\pi, \pi]$ and phase $\beta \in [-\pi, \pi]$
of the form (see Figure~\ref{fig-schem-ill-model-rf-sine})
\begin{equation}
  \label{eq-sine-wave-model-spat-anal}
  f(x_1, x_2) =
  \sin
  \left(
    \omega \cos (\theta) \, x_1 + \omega \sin (\theta) \, x_2+ \beta
  \right).
\end{equation}
Additionally, the responses of the above joint spatio-temporal models of receptive fields are calculated with respect to a moving sine wave of the form
\begin{multline}
  \label{eq-moving-sine-wave}
  f(x_1, x_2, t) = \\ 
  = \sin
     \left(
       \omega \cos (\theta) \, (x_1 - u_1 t) + \omega \sin (\theta) \, (x_2 -  u_2 t) + \beta
     \right),
\end{multline}
with  the velocity vector $(u_1, u_2)^T$ parallel to the inclination angle
$\theta$ of the grating, such that $(u_1, u_2)^T = (u \cos \theta, u
\sin \theta)^T$ for $u = \sqrt{u_1^2 + u_2^2}$.

In summary, the theoretical analysis in  Lindeberg (\citeyear{Lin25-JCompNeurSci-orisel})
shows that the resulting orientation selectivity curves for the first-order simple cells, second-order simple cells and complex cells, respectively, will be of the forms:
\begin{align}
  \begin{split}
    \label{eq-ori-sel-r-simple1}
    r_{\simple,1}(\theta)
    & = \frac{\left| \cos \theta \right|}
                   {\sqrt{\cos ^2 \theta + \kappa ^2 \sin ^2\theta}},
  \end{split}\\
  \begin{split}
    \label{eq-ori-sel-r-simple2}    
    r_{\simple,2}(\theta)
    & = \frac{\cos^2 \theta}
                   {\cos ^2 \theta + \kappa ^2 \sin ^2\theta},
  \end{split}\\
  \begin{split}
    \label{eq-ori-sel-r-complex}    
    r_{\complex}(\theta)
    & = \frac{\left| \cos \theta \right|^{3/2}}
                      {\left( \cos ^2 \theta + \kappa ^2 \sin ^2\theta \right)^{3/4}},
  \end{split}
\end{align}
with similar angular dependencies within each class for both the
purely spatial receptive fields and the joint spatio-temporal
receptive fields,
where
\begin{equation}
  \kappa = \frac{\sigma_2}{\sigma_1}
\end{equation}
denotes the ratio between the scale parameters $\sigma_2 \in \bbbr_+$ and
$\sigma_1 \in \bbbr_+$ in the vertical and horizontal directions of the
affine Gaussian kernel that determines the spatial shape of
the receptive field.

The top row in Figure~\ref{fig-ori-sel} shows graphs of these orientation selectivity curves, where we can clearly see how the orientation selectivity becomes more narrow for increasing values of the scale parameter ratio $\kappa$, thus establishing a direct link between the elongation and the degree of orientation selectivity for these idealized receptive fields.

In Appendix~\ref{sec-ori-sel-order-3-4},
we additionally extend these results to
orientation selectivity for purely spatial third-order simple cells
and fourth-order simple cells of the forms
\begin{align}
  \begin{split}
    \label{eq-ori-sel-r-simple3}
    r_{\simple,3}(\theta)
    & = \frac{\left| \cos \theta \right|^3}
                   {(\cos ^2 \theta + \kappa ^2 \sin ^2\theta)^{3/2}},
  \end{split}\\
  \begin{split}
    \label{eq-ori-sel-r-simple4}    
    r_{\simple,4}(\theta)
    & = \frac{\cos^4 \theta}
                   {(\cos ^2 \theta + \kappa ^2 \sin ^2\theta)^2},
  \end{split}
\end{align}
and with examples of graphs of these curves shown in
the bottom row in Figure~\ref{fig-ori-sel}
for a few values of the scale ratio parameter $\kappa$.
Also for the third- and fourth-order receptive fields, 
the orientation selectivity curves become more narrow, both with
increasing values of the scale parameter ratio $\kappa$ and with
increasing order of spatial differentiation.

\section{Modelling the orientation selectivity properties of simple
and complex cells in the primary visual cortex in terms of affine
Gaussian derivative based receptive fields with a variability in eccentricity}
\label{sec-results}

\subsection{Interpretation of the connection between the orientation selectivity and the elongation of receptive fields in relation to biological measurements}
\label{sec-rel-biol-vision}

In this section, we will compare the results of the theoretical
predictions in Section~\ref{sec-conn-ori-sel-elong}
with biological results concerning the orientation
selectivity of  visual neurons.

\subsubsection{Interpretation of the measurements about broad vs.\ sharp
  orientation selectivity of neurons by Nauhaus et al. (\citeyear{NauBenCarRin09-Neuron})}
\label{sec-interpret-nauhaus-exps}

\begin{figure}[hbtp]
  \begin{center}
    \includegraphics[width=0.20\textwidth]{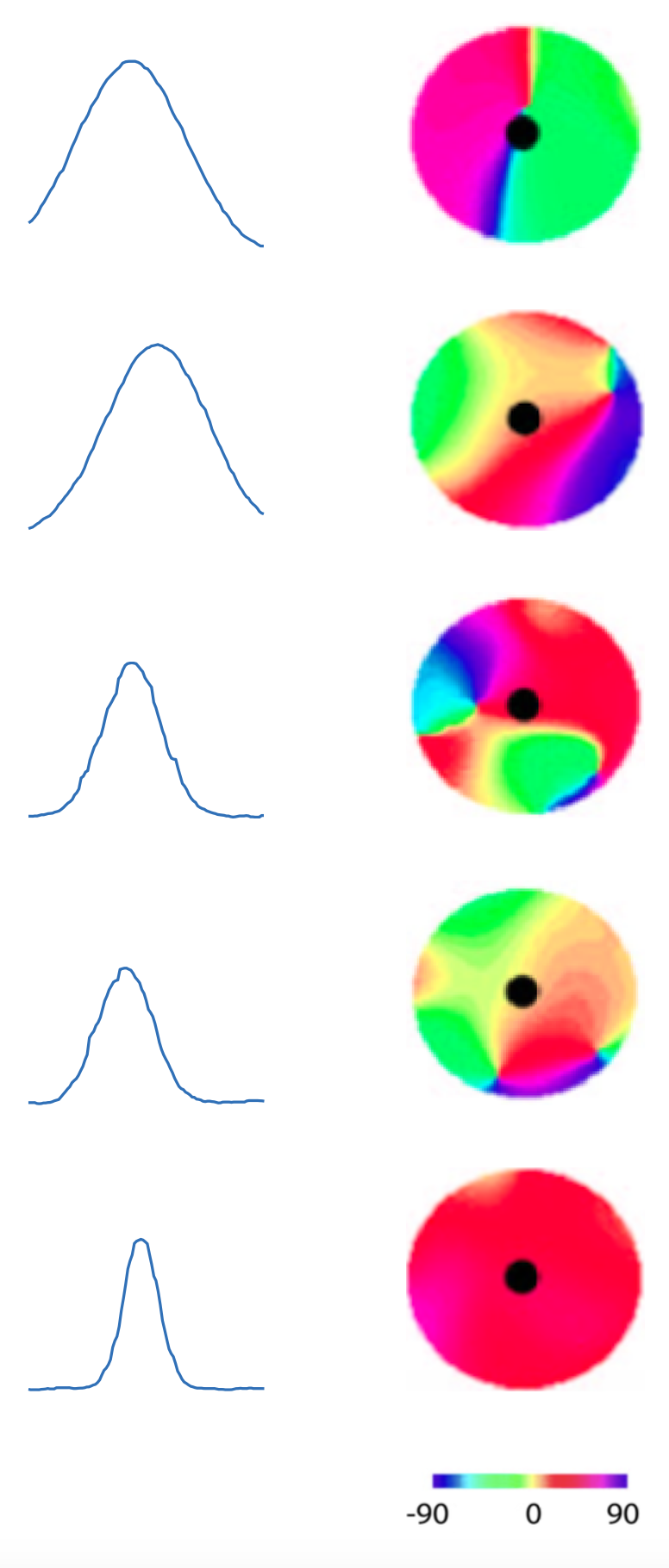}
  \end{center}
  \caption{Schematic depiction of results from measurements of the orientation tuning of neurons, at
    different positions in the visual cortex, adapted from
    Nauhaus {\em et al.\/}\ (\citeyear{NauBenCarRin09-Neuron}),
    showing how the orientation tuning
    changes from broad to sharp, and thus higher degree of orientation
  selectivity, with increasing distance from the pinwheels, consistent
with the qualitative behaviour that would be obtained if the ratio
$\kappa$, between the scale parameters in underlying affine Gaussian
smoothing step in the idealized models of spatial and spatio-temporal
receptive fields, would increase when moving away from the centers of
the pinwheels on the cortical surface.}
  \label{fig-nauhaus-ori-tun-meas}
\end{figure}

Nauhaus {\em et al.\/}\ (\citeyear{NauBenCarRin09-Neuron}) have
measured the orientation tuning of neurons at different positions
in the primary visual cortex for monkey and cat. They found that the
orientation tuning is broader near the pinwheel centers and sharper
in regions of homogeneous orientation preference, 
see specifically Figure~2 in Nauhaus {\em et al.\/}\ (\citeyear{NauBenCarRin09-Neuron}).
Figure~\ref{fig-nauhaus-ori-tun-meas} shows a schematic depiction of their
results, where the degree of orientation selectivity
changes%
\footnote{In Figure~\ref{fig-nauhaus-ori-tun-meas},
  the orientation selectivity is widest in the top row, corresponding to
the center of the pinwheel. The orientation selectivity is most
narrow in the bottom row, corresponding to a position in the visual
cortex far away from the center of the pinwheel. Then, for the three
intermediate rows, the orientation selectivity is intermediate, and
corresponding to intermediate positions in the visual cortex some
distance away from the centers of the pinwheels.}
from broad to sharp with increasing distance from
the pinwheel center (from top to bottom in the figure),
however, more clearly visible in the original Figure~2 in
Nauhaus {\em et al.\/}\ (\citeyear{NauBenCarRin09-Neuron}).
This behaviour is also in agreement with the observation by
Wilson {\em et al.\/} (\citeyear{WilWhiSchFit16-NatNeuroSci})
that ``neurons located near pinwheel centers
... exhibit broader orientation tuning than neurons in regions of the
map where neighbouring neurons exhibit similar preferences''.

Thus, first of all, we can see that the variability in the
orientation selectivity of visual neurons is consistent with a
variability in the degree of elongation of the receptive fields, as
predicted by the underlying normative theory of visual receptive
fields, and also consistent with the more general hypotheses, that the family
of receptive fields ought to be expanded over the degrees of freedom
of affine image transformations to be affine covariant.

Secondly,
this qualitative behaviour is consistent with what would be the result
if the ratio $\kappa$ between the two scale parameters of
the underlying affine Gaussian kernels
would increase from a lower to a higher value,
when moving away from the centers of the pinwheels on the cortical surface.
Thus, by combination with the experimental results by
Nauhaus {\em et al.\/}\ (\citeyear{NauBenCarRin09-Neuron})
and Wilson {\em et al.\/} (\citeyear{WilWhiSchFit16-NatNeuroSci}),
the presented theory leads to a prediction about a systematic variability
in the eccentricity or the elongation
of the receptive fields in the primary visual cortex,
which for the case of pinwheel structures, would
be consistent with a variability in the eccentricity or
the elongation of the receptive fields 
from the centers of the pinwheels towards the periphery.


A highly interesting quantitative measurement to perform, in view of
these theoretical results, would hence
be to fit parameterized models of the orientation selectivity,
according to Equations~(\ref{eq-ori-sel-r-simple1}),
(\ref{eq-ori-sel-r-simple2}), (\ref{eq-ori-sel-r-complex}),
(\ref{eq-ori-sel-r-simple3}) and (\ref{eq-ori-sel-r-simple4})
\begin{equation}
  \label{eq-ori-sel-curves-gen-gauss-der-model-general}
  r_{\lambda}(\theta)
  = \left(
         \frac{\left| \cos \theta \right|}
         {\sqrt{\cos ^2 \theta + \kappa ^2 \sin ^2\theta}}
         \right)^{\lambda}
 \end{equation}
for $\lambda \in \bbbr$ to orientation tuning curves of the form recorded by
Nauhaus {\em et al.\/}\ (\citeyear{NauBenCarRin09-Neuron}), to get
estimates of the distribution of the parameter $\kappa$ over a
sufficiently large population of visual neurons, under the assumption that the
spatial components of the biological receptive fields can be well
modelled by affine Gaussian derivatives.%
\footnote{At the point of writing this
article, the author does, however, not have access to the explicit raw
data that would be needed to perform such an analysis.}


If we would assume that it would be unlikely for the receptive fields to have as
strong variability in their orientational selectivity properties as a
function of the positions of the neurons in relation to the pinwheel
structure, as reported in this study, without also having a strong variability in their
eccentricity. Then, by combining the theoretical analysis in this article with the
biological results by Nauhaus {\em et al.\/}\ (\citeyear{NauBenCarRin09-Neuron}),
that would
serve as possible indirect support for the hypothesis concerning an expansion
of receptive field shapes over variations in the ratio between the two scale parameters
of spatially anisotropic receptive fields.


\begin{figure*}[!h]
  \begin{center}
    \begin{tabular}{cc}
      {\em\small Simple cells (n = 142)\/}
      &  {\em\small Complex cells (n = 184)\/} \\
      \includegraphics[width=0.30\textwidth]{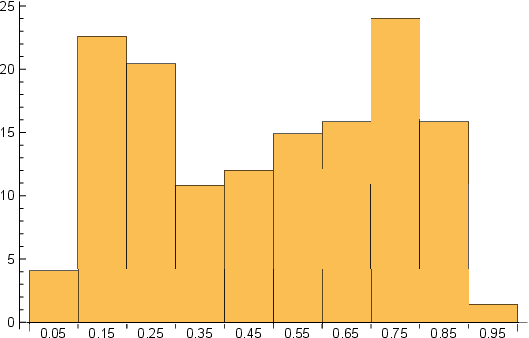}
      & \includegraphics[width=0.30\textwidth]{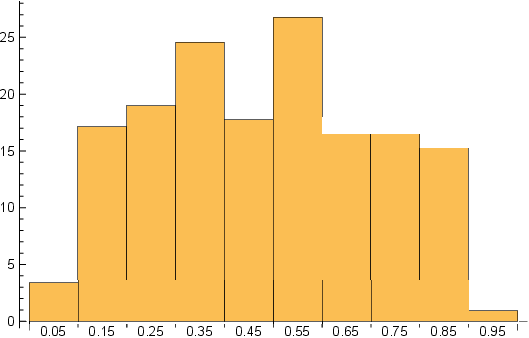}
     \end{tabular}
   \end{center}
   \caption{Schematic depictions of the shapes of the distributions of the absolute value of the resultant
     $|R| \in [0, 1]$ according to (\ref{eq-def-resultant}) for the
     directional selectivity of visual neurons over populations
     of simple cells and complex cells, respectively, in the primary visual cortex,
     adapted from neurophysiological recordings of Macaque monkeys by
     Goris {\em et al.\/}  (\citeyear{GorSimMov15-Neuron}).
     (Horizontal axes: 10 quantized bins over the resultant $|R| \in [0, 1]$.
     Vertical axes: approximate bin counts.)}
   \label{fig-goris-orient-select-stat}
\end{figure*}

\subsubsection{Qualitative interpretation of the measured orientation
  selectivity histograms by Goris et al.\ (\citeyear{GorSimMov15-Neuron})}
\label{sec-interpret-goris-exps-1-2}

These predictions are furthermore consistent with existing biological results by
Goris {\em et al.\/} (\citeyear{GorSimMov15-Neuron}), concerning the
distribution of the degree of orientation selectivity of the
neurons in the primary visual cortex.
By measuring the absolute value $|R| \in \bbbr$ of the
complex-valued resultant, given by
\begin{equation}
  \label{eq-def-resultant}
  R = \frac{\int_{\theta = - \pi}^{\pi} r(\theta) \, e^{2 i \theta} d\theta}
                {\int_{\theta = - \pi}^{\pi} r(\theta) \, d\theta},
\end{equation}
for each visual neuron, and then computing a normalized histogram of
these measurements (see Figure~\ref{fig-goris-orient-select-stat}
for a schematic depiction of the results in Figure~1B in
Goris {\em et al.\/} (\citeyear{GorSimMov15-Neuron})),
Goris {\em et al.\/} (\citeyear{GorSimMov15-Neuron})
demonstrate a substantial variability in the orientation selectivity
of the receptive fields of simple cells and complex cells in the
primary visual cortex.

\begin{figure*}[tbp]
  \begin{equation}
    R_{\complex}
    = \frac{\kappa ^2 \left(48 \left(\kappa ^2-1\right)^{3/4} \Gamma \left(\frac{5}{4}\right)
   \, _2F_1\left(\frac{1}{2},1;\frac{3}{4};\frac{1}{\kappa ^2}\right)-16 \left(\kappa
   ^2-1\right)^{3/4} \Gamma \left(\frac{5}{4}\right)+3 \sqrt{2 \pi } \, \kappa  \, \Gamma
   \left(-\frac{1}{4}\right)\right)}{2 \left(\kappa ^2-1\right) \left(16 \left(\kappa
   ^2-1\right)^{3/4} \Gamma \left(\frac{5}{4}\right) \,
   _2F_1\left(\frac{1}{2},1;\frac{3}{4};\frac{1}{\kappa ^2}\right)+\sqrt{2 \pi } \, \kappa \,
   \Gamma \left(-\frac{1}{4}\right)\right)}
  \end{equation}
  \caption{Closed-form expression for the resultant $R$ according
    to (\ref{eq-def-resultant}) calculated for
    the orientation selectivity curves (\ref{eq-ori-sel-r-complex})
    for our idealized models of
    complex cells, valid for the purely spatial model
    (\ref{eq-quasi-quad-dir}) and the joint
    spatio-temporal model
    (\ref{eq-quasi-quad-dir-vel-adapt-spat-temp}).
    The function $_2F_1(a, b; c; z)$ denotes the hypergeometric
    function $\operatorname{Hypergeometric2F1}[a, b, c, z]$ in
    Mathematica, while $\Gamma(z)$  represents Euler's Gamma function.}
  \label{fig-expl-expr-R-complex-2-cases}
\end{figure*}

\begin{figure}[hbtp]
  \begin{center}
    \includegraphics[width=0.45\textwidth]{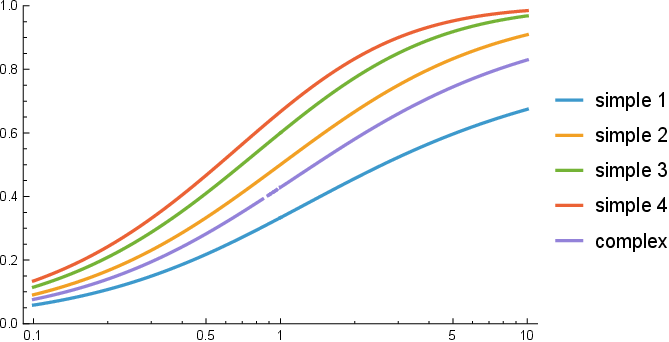}
  \end{center}
  \caption{Graphs of the resultant $R$ for idealized models of
    (i) a first-order simple cell according to
    (\ref{eq-resultant-1st-order-simple}),
    (ii) a second-order simple cell according to
    (\ref{eq-resultant-2nd-order-simple}),
    (iii) a third-order simple cell according to
    (\ref{eq-resultant-3rd-order-simple}),
    (iv) a fourth-order simple cell according to
    (\ref{eq-resultant-4th-order-simple})  and
    (v) a complex cell according to (\ref{eq-resultant-complex})
    on a log-linear scale, with the horizontal axis parameterized in
    terms of the logarithm $K = \log \kappa$ of the scale parameter
    ratio $\kappa$ in the affine Gaussian derivative model of visual
    receptive fields. (Horizontal axes: scale parameter
    ratio~$\kappa$. Vertical axes: resultant~$R$.)}
  \label{fig-graphs-resultant-loglinear}
\end{figure}

This consistency can be demonstrated
by computing the closed-form expression for the absolute
value of the resultant of the orientation selectivity curves
according to
Equations~(\ref{eq-ori-sel-r-simple1})--(\ref{eq-ori-sel-r-complex})
and Equations~(\ref{eq-ori-sel-r-simple3})--(\ref{eq-ori-sel-r-simple4}):
\begin{itemize}
\item
  for the first-order idealized models of simple cells
\begin{align}
   \begin{split}
      R_{\simple,1}
      & = \frac{\int_{\theta = -\pi/2}^{\pi/2}
                        \frac{ \cos \theta }
                                {\sqrt{\cos ^2 \theta + \kappa ^2 \sin ^2\theta}}
                        \, \cos 2 \theta \, d\theta}
                     {\int_{\theta = -\pi/2}^{\pi/2}
                         \frac{ \cos \theta}
                                 {\sqrt{\cos ^2 \theta + \kappa ^2 \sin ^2\theta}}
                         \, d\theta}
    \end{split}\nonumber\\
    \begin{split}
      \label{eq-resultant-1st-order-simple}
      & = \frac{\kappa  \left(\kappa  \cosh ^{-1} \kappa -\sqrt{\kappa ^2-1}\right)}{\left(\kappa
   ^2-1\right) \cosh ^{-1} \kappa},
     \end{split}
\end{align}
\item
  for the second-order idealized models of simple cells
\begin{align}
   \begin{split}
     R_{\simple,2}
     & = \frac{\int_{\theta = -\pi/2}^{\pi/2}
                         \frac{\cos^2 \theta}
                                 {\cos ^2 \theta + \kappa ^2 \sin ^2\theta}
                         \, \cos 2 \theta \, d\theta}
                    {\int_{\theta = -\pi/2}^{\pi/2}
                         \frac{\cos^2 \theta}
                                 {\cos ^2 \theta + \kappa ^2 \sin ^2\theta}
                         \, d\theta}
    \end{split}\nonumber\\
    \begin{split}
      \label{eq-resultant-2nd-order-simple}
       & = \frac{\kappa }{\kappa +1},
    \end{split}
\end{align}
\item
  for the idealized models of complex cells
\begin{align}
  \begin{split}
    \label{eq-resultant-complex}
     R_{\complex}
     & = \frac{\int_{\theta = -\pi/2}^{\pi/2}
                        \frac{\cos ^{3/2} \theta }
                                {\left( \cos ^2 \theta + \kappa ^2 \sin ^2\theta \right)^{3/4}}
                                \, \cos 2 \theta \, d\theta}
                    {\int_{\theta = -\pi/2}^{\pi/2}
                        \frac{\cos ^{3/2} \theta}
                                {\left( \cos ^2 \theta + \kappa ^2 \sin ^2\theta \right)^{3/4}}
                                \, d\theta},               
    \end{split}
\end{align}
with the explicit expression for that result in
Figure~\ref{fig-expl-expr-R-complex-2-cases},
\item
  for the third-order idealized models of simple cells
\begin{align}
   \begin{split}
      R_{\simple,3}
      & = \frac{\int_{\theta = -\pi/2}^{\pi/2}
                        \frac{ \cos^3 \theta }
                                {(\cos ^2 \theta + \kappa ^2 \sin ^2\theta)^{3/2}}
                        \, \cos 2 \theta \, d\theta}
                     {\int_{\theta = -\pi/2}^{\pi/2}
                         \frac{ \cos^3 \theta}
                                 {(\cos ^2 \theta + \kappa ^2 \sin ^2\theta)^{3/2}}
                         \, d\theta}
    \end{split}\nonumber\\
    \begin{split}
      \label{eq-resultant-3rd-order-simple}
      & = \frac{\kappa  \left(\sqrt{\kappa ^2-1} \left(\kappa ^2+2\right)-3 \kappa  \cosh
   ^{-1}(\kappa )\right)}{\left(\kappa ^2-1\right) \left(\kappa  \sqrt{\kappa
   ^2-1}-\cosh ^{-1}(\kappa )\right)},
     \end{split}
\end{align}
\item
  and for the fourth-order idealized models of simple cells
\begin{align}
   \begin{split}
     R_{\simple,4}
     & = \frac{\int_{\theta = -\pi/2}^{\pi/2}
                         \frac{\cos^4 \theta}
                                 {(\cos ^2 \theta + \kappa ^2 \sin ^2\theta)^2}
                         \, \cos 2 \theta \, d\theta}
                    {\int_{\theta = -\pi/2}^{\pi/2}
                         \frac{\cos^4 \theta}
                                 {(\cos ^2 \theta + \kappa ^2 \sin ^2\theta)^2}
                         \, d\theta}
    \end{split}\nonumber\\
    \begin{split}
      \label{eq-resultant-4th-order-simple}
       & = \frac{\kappa  \, (\kappa +3)}{(\kappa +1) (\kappa +2)}.
    \end{split}
\end{align}
\end{itemize}
Let us additionally reparameterize these curves in terms of
a logarithmic parameterization $K = \log \kappa$ of the scale parameter ratio
$\kappa$, which leads to graphs 
shown in Figure~\ref{fig-graphs-resultant-loglinear}.
Then, we can see that the experimentally obtained
distributions in Figure~\ref{fig-goris-orient-select-stat}
appear to be reasonably consistent%
\footnote{If the logarithmic parameterization of the scale parameter
  ratio $\kappa$ would be a perfect prior for resultant measure $R$,
  then these graphs would be pure linear functions, so that the
  corresponding histograms over the resultant $R$ would then be constant.}
with the assumption of a rather uniform distribution over the
logarithmically parameterized scale parameter ratio $K = \log \kappa$.

Such a parameterization would specifically constitute a canonical parameterization, if one would
simplify%
\footnote{\label{footnote-hemisphere-distr}
  More generally, one could instead conceive a uniform joint
  distribution on a hemisphere, as conceived in Figure~8
  in (Lindeberg \citeyear{Lin21-Heliyon}), which regarding first-order
  spatial derivatives then leads to a distribution of spatial receptive field
  shapes of the form shown in Figure~\ref{fig-aff-gauss-1der-distr},
  and possibly complemented with additional priors to account for
  how important different local surface orientations in the
  environment would be for the perceptual process, as well as
  densely the space of combined image orientations $\varphi$
  and scale parameter ratios $\kappa$ would need to be sampled,
  to support sufficiently good approximations of covariance
  over that submanifold for the local image measurements performed
  by the family of spatial receptive fields.
  In this treatment, we do, however, simplify this problem, by instead
  considering a uniform distribution over a logarithmic transformation
  of the scale parameter ratio $\kappa$, which is also easier to handle
  in closed form calculations, and which may be regarded as a
  coarse approximation, to compensate for gross phenomena with
  regard to a non-uniform distribution of receptive field shapes
  over the scale parameter ratio $\kappa$.}
the 2-D joint distribution of receptive field shapes over
the scale parameter ratio $\kappa$ and the orientation $\varphi$ into two
independent 1-D distributions over the scale parameter ratio $\kappa$
and the orientation $\varphi$, respectively, in the idealized
model of visual receptive fields according to the generalized Gaussian
derivative framework.

Thus, also these biological results are qualitatively
consistent with the working hypothesis about an expansion over 
the degree of elongation of the receptive fields in the primary visual
cortex, as would be implied from the assumption of a family of
affine covariant visual receptive fields.

\subsubsection{Quantitative modelling of the measured orientation
  selectivity histograms by Goris et al.\ (\citeyear{GorSimMov15-Neuron})}
\label{sec-interpret-goris-exps-3-4}

To aim at more detailed {\em quantitative\/} modelling of the
experimentally recorded histograms of the resultant measure of the
orientation selectivity curves, as reported by Goris {\em et al.\/}
(\citeyear{GorSimMov15-Neuron}) and as schematically reproduced in 
Figure~\ref{fig-goris-orient-select-stat}, we need to
consider that there are more free parameters
in the modelling stage to determine, based on the following arguments:
\begin{itemize}
\item
  One basic question concerns what range of values of the scale
  parameter ratio $\kappa$ would be spanned by the receptive fields in
  the primary visual cortex.%
\footnote{In the graphs shown in 
  Figure~\ref{fig-graphs-resultant-loglinear}, we have used a range of
  the scale parameter ratio $\kappa$
  over a factor 10 from the unit value 1, which results in a maximum
  value of $R$ for the idealized first-order model of a simple cell 
  of about 0.67. In contrast,
  for the idealized second-order model of a simple cell,
  $R$ reaches a maximum value of $R$ about 0.91,
  while for the complex cell,
  $R$ assumes a maximum value of about 0.83.
  For the third-order simple cell the maximum value of $R$ is about 0.97
  and for the fourth-order simple cell the maximum value of $R$ is
  about 0.98.
  If this range of scale parameter ratios would be reduced to a
  lower span, then the range of the possible values of $R$ would be
  reduced, while the range would be expanded if a wider range of scale
  ratios $\kappa$ would be implemented.}
\item
  Another basic question concerns the distribution of receptive fields
  with respect to the order of spatial differentiation.
%
%
%
  To reproduce an idealized model of a histogram of the resultant $R$
  for a population of simple cells, as shown in the top part of
  Figure~\ref{fig-goris-orient-select-stat}, we would therefore
  have to a assume a distribution of receptive fields over different
  orders of spatial differentiation.
\end{itemize}

\begin{figure*}[hbtp]
  \begin{center}
    \begin{tabular}{ccc}
      {\em\footnotesize First-order simple cells\/}
      &       {\em\footnotesize Second-order simple cells\/}
      &       {\em\footnotesize Complex cells\/} \\
      \includegraphics[width=0.30\textwidth]{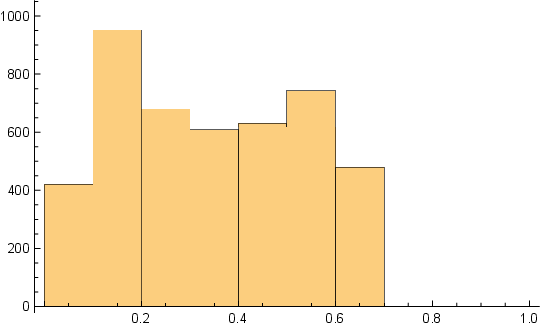}
      & \includegraphics[width=0.30\textwidth]{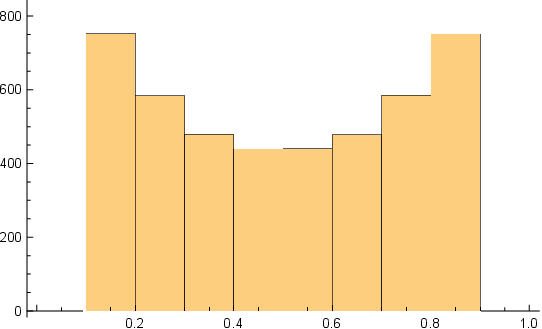}
      & \includegraphics[width=0.30\textwidth]{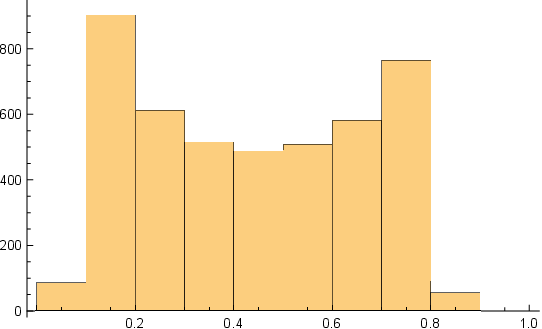}\\
      \\
      {\em\footnotesize Third-order simple cells\/}
      &       {\em\footnotesize Fourth-order simple cells\/} \\
      \includegraphics[width=0.30\textwidth]{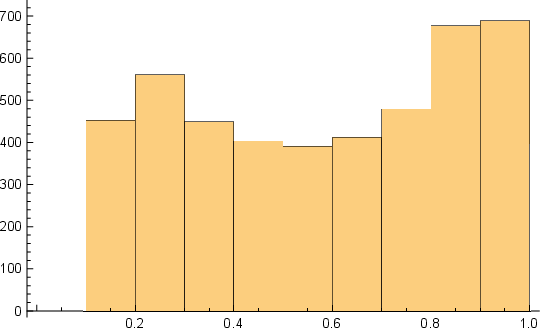}
      & \includegraphics[width=0.30\textwidth]{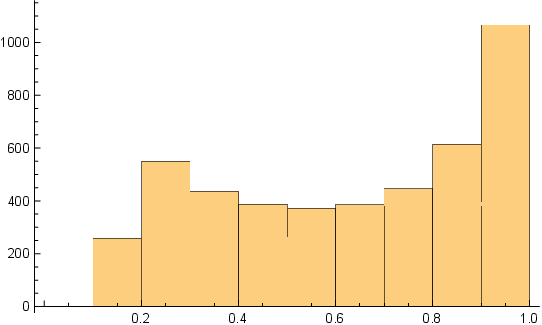} \\    
    \end{tabular}
  \end{center}
  \caption{Examples of histograms of the resultant $R$ over
    populations of (top left) first-order simple cells, (top middle)
    second-order simple cells, (top right) complex cells,
    (bottom left) third-order simple cells and (bottom right)
    fourth-order simple cells, accumulated
    over a uniform logarithmic distribution of the scale parameter
    ratio $\kappa$ over the interval $\kappa \in [1/\kappa_{\max}, \kappa_{\max}]$
    for $\kappa_{\max} = 8$.
  (Horizontal axes: bin over the resultant $R$. Vertical axes: number of receptive
  fields in this bin in a discrete simulation.)}
  \label{fig-histograms-resultant}
\end{figure*}

\begin{figure*}[hbtp]
  \begin{center}
    \begin{tabular}{cc}
      {\em\footnotesize Simple cells up to order 2\/}
      &       {\em\footnotesize Simple cells up to order 4\/} \\
      \includegraphics[width=0.30\textwidth]{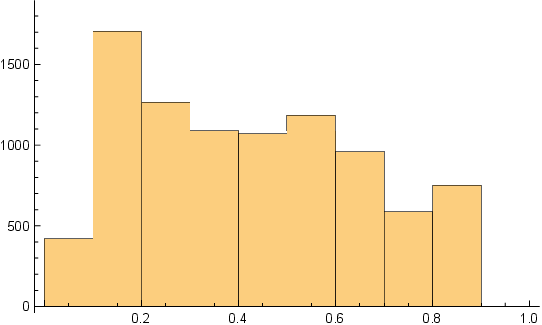}
      & \includegraphics[width=0.30\textwidth]{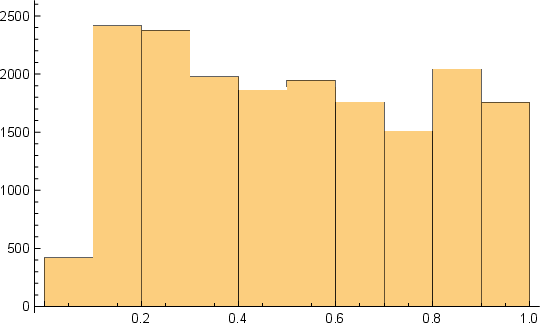} \\
    \end{tabular}
  \end{center}
  \caption{Examples of combined histograms of the resultant $R$ over
    populations of simple cells of different order (left) up to order 2 and (right)
    up to order 4, accumulated
    over a uniform logarithmic distribution of the scale parameter
    ratio $\kappa$ over the interval $\kappa \in [1/\kappa_{\max}, \kappa_{\max}]$
    for $\kappa_{\max} = 8$.
    (Horizontal axes: bin over the resultant $R$. Vertical axes: number of receptive
  fields in this bin in a discrete simulation.)}
  \label{fig-histograms-resultant-combined}
\end{figure*}


\noindent
To illustrate to what extent the distributions of the resultant will
be influenced by receptive fields for different orders of spatial
differentiation, 
Figure~\ref{fig-histograms-resultant} shows
histograms of the resultant accumulated for idealized models of simple
cells of orders 1, 2, 3 and 4 as well as for the idealized model of complex
cells.
For generating these graphs, we have created a uniform distribution of
the scale parameter ratio $\kappa$ over a logarithmic scale, over the
interval $\kappa \in[1/\kappa_{\max}, \kappa_{\max}]$ for the
arbitrary choice of $\kappa_{\max} = 8$.

As can be seen from these graphs, the histogram over the first-order
simple cells is delimited by a maximum value around
$R_{\max,1} \approx 0.7$, while the distributions for third-order and
fourth-order simple cells are heavier for larger values of the
resultant approaching $R \rightarrow 1$.
The model used for computing histograms of the resultant for the
idealized models of complex cells does, however, not very well reproduce
the shape of the biologically obtained histogram, thus indicating that the model
for the complex cells may be overly simplified%
\footnote{In this context, it should be remarked that there is strong
  conceptual difference between the idealized models for simple cells
  {\em vs.\/} the idealized models of complex cells. The idealized model
for simple cells has been determined in a theoretically principled
manner from axiomatic derivations, and also been matched to biological
measurements of simple cells, whereas the idealized model for complex
cells has been chosen as an as straightforward way as possible for
combining the responses of odd-shaped and even-shaped simple cells.}
for the purpose of
reproducing the shape of the resultant histogram. See the discussion
in Section~6 in  (Lindeberg \citeyear{Lin25-JCompNeurSci-orisel})
for a number of suggested ways to extend that model.

Figure~\ref{fig-histograms-resultant-combined} shows additional
results of combining the resultant for populations of simple cells
over different orders of spatial differentiation, either up to order~2
or up to order~4, here assuming the same number of neurons
for all the different orders of spatial differentiation.

With ample reservation from the fact that this theoretical analysis is
conceptually simplified in a number of ways,%
\footnote{The reservation that we have to state concerning this theoretical modelling of
the resultant histograms accumulated by by Goris {\em et al.\/}
(\citeyear{GorSimMov15-Neuron}), however, is that the analysis is
simplified in the following 5 major ways:
(i)~In the current state, we
  do not have any principled biological arguments for choosing particular
  values of the parameters $\kappa_{\min}$ and $\kappa_{\max}$ that
  determine the range of the scale parameter ratio $\kappa$,
  where different choices of these parameters may affect the shapes
  of the combined histograms of the resultant $R$.
(ii)~The choice of a logarithmic distribution over the scale parameter
  $\kappa$ does, as previously mentioned, neglect any co-dependency with
  respect to the distribution over the orientation angle $\varphi$.
(iii)~The experimental conditions used when accumulating the orientation
  selectivity curves for biological neurons may not be identical to
  the conditions under which the theoretically derived orientation
  selectivity curves have been derived.
(iv)~The assumption about equal numbers of receptive fields for the
  different orders of spatial differentiation may not necessarily hold
  in reality.
(v)~Our computations of the
  resultant $R$ for the receptive fields are based on an idealized noise
  free model, while there additionally could be sources of noise in the
  biological experiments as well as modelling errors between the
  receptive fields of the actual biological neurons and our
  idealized receptive fields.}
as can be seen from these results, the combined histograms give rise
to a bump in the histograms for lower values of the resultant $R$,
in qualitative agreement with the biological results by Goris {\em et al.\/}
(\citeyear{GorSimMov15-Neuron}) and as reproduced in 
Figure~\ref{fig-goris-orient-select-stat}.
Furthermore, to obtain something that
would look like a small bump for larger values of the resultant $R$,
the modelling situation with receptive fields up to order~4 
gives a closer similarity to the biological results by Goris {\em et al.\/}
(\citeyear{GorSimMov15-Neuron}) compared to the model based on receptive
fields up to order~2.

Assuming that the modelling simplifications do not significantly affect the
qualitative nature of the results, from the presented analysis it thus appears as if:
\begin{itemize}
\item
  The histograms of the
resultant of the simple cells in the biological experiments
{\em et al.\/} (\citeyear{GorSimMov15-Neuron}) could be rather well
explained by the receptive fields in the primary
visual cortex of Macaque monkeys having a variability over the degree
of elongation.
\item
  The
non-uniform nature of the experimentally obtained histogram of the resultant $R$ for the
simple cells could be explained better by assuming that receptive fields
should be present up to a spatial differentiation order up to 4 than
up to a spatial differentiation order of 2.
\end{itemize}

\subsubsection{Summary of the interpretations and modelling results of the biological
experiments by Nauhaus et al.\ (\citeyear{NauBenCarRin09-Neuron}) and 
by Goris et al.\ (\citeyear{GorSimMov15-Neuron})}

To conclude, the biological results by
Nauhaus {\em et al.\/}\ (\citeyear{NauBenCarRin09-Neuron}),
Goris {\em et al.\/} (\citeyear{GorSimMov15-Neuron}) and
Wilson {\em et al.\/} (\citeyear{WilWhiSchFit16-NatNeuroSci})
are clearly
consistent with an expansion over the degree of elongation
of the receptive field shapes in the primary visual cortex.

Based on these results we propose that, beyond an expansion over
rotations, as is performed in current models of the pinwheel structure
of visual receptive fields
(Bonhoeffer and Grinvald \citeyear{BonGri91-Nature},
Blasdel \citeyear{Bla92-JNeuroSci},
Swindale \citeyear{Swi96-NetwCompNeurSys},
Petitot \citeyear{Pet03-JPhysPar},
Koch {\em et al.\/} \citeyear{KocJinAloZai16-NatComm},
Kremkow {\em et al.\/}\ \citeyear{KreJinWanAlo16-Nature},
 Baspinar {\em et al.\/}\ \citeyear{BasCitSar18-JMIV},
 Najafian {\em et al.\/}\ \citeyear{NajKocTehJinRahZaiKreAlo22-NatureComm},
 Liu and Robinson \citeyear{LiuRob22-FronCompNeurSci}),
also an explicit expansion over the eccentricity $\epsilon$ of the
receptive fields (the inverse of the parameter $\kappa$) should be
included, when modelling the pinwheel structure in the visual cortex.

Possible ways, by which an explicit dependency on the eccentricity of
the receptive fields could
be incorporated into the modelling of pinwheel structures, will be
outlined in more detail in the following treatment regarding more specific
biological hypotheses.

\subsection{Explicit testable hypotheses for biological experiments}
\label{sec-biol-hypo}

Based on the above theoretical analysis, with its associated theoretical
predictions, we propose that it would
be highly interesting to perform experimental characterization and
analysis based on joint estimation of
\begin{itemize}
\item
  orientational selectivity,
\item
  receptive field eccentricity,
\item
  orientational homogeneity and
\item
  location of the neuron in the visual cortex in relation to the pinwheel
  structure,
\end{itemize}
in the
primary visual cortex of animals with clear pinwheel structures, to
determine if there is a variability in the eccentricity or elongation of the receptive
fields, and specifically if the degree of elongation increases with
the distance from the centres of the pinwheels towards periphery,
as arising as one possible interpretation of combining the theoretical
results about orientation selectivity of affine Gaussian receptive
fields in this article with the biological results by
Nauhaus {\em et al.\/}\ (\citeyear{NauBenCarRin09-Neuron}).

\begin{figure}[hbtp]
   \begin{center}
    \begin{tabular}{c}
     \includegraphics[width=0.35\textwidth]{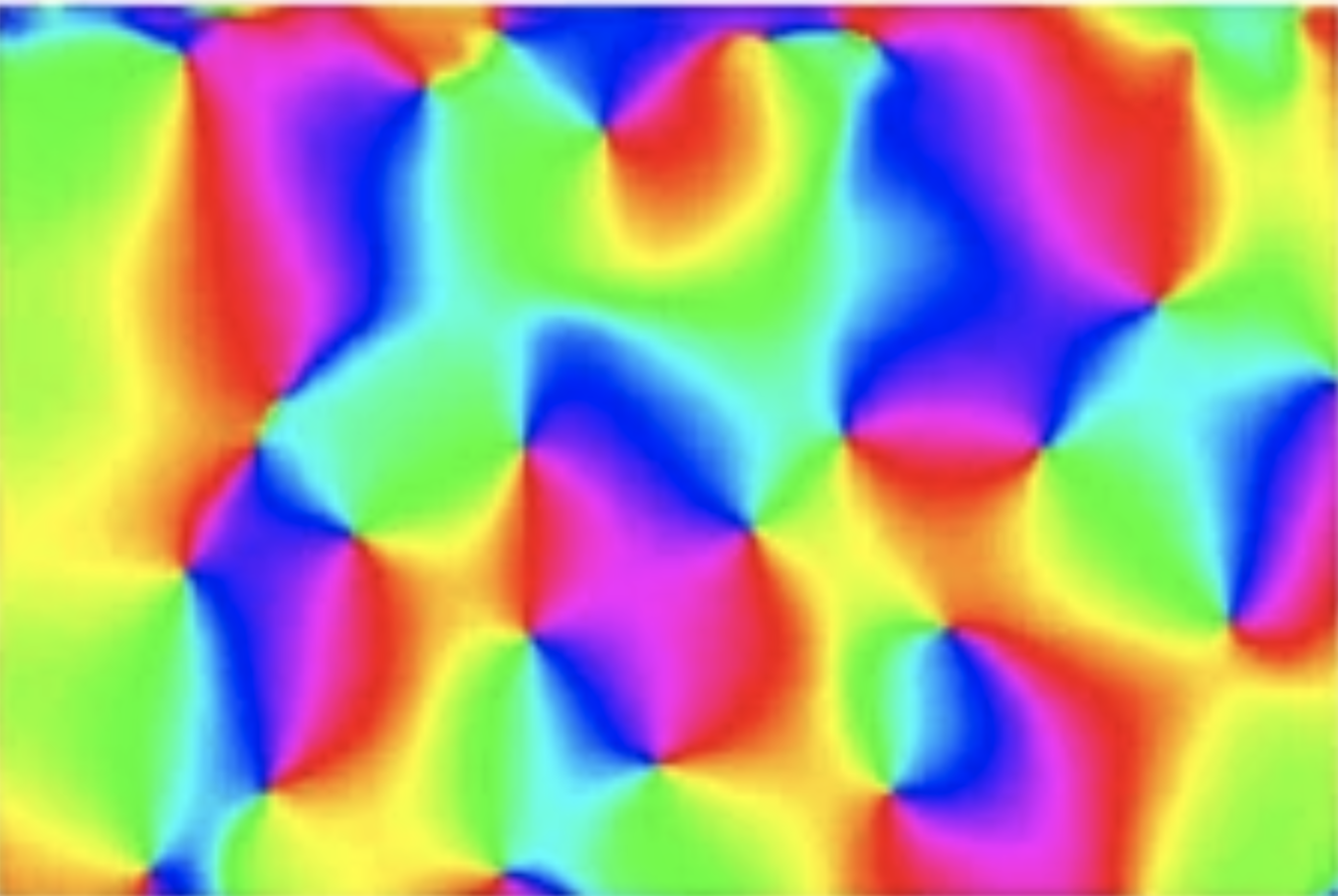}
    \end{tabular}
    \end{center}
    \caption{Orientation map in the primary visual cortex of cat, 
     as recorded by Koch {\em et al.\/} (\citeyear{KocJinAloZai16-NatComm})
     (OpenAccess), with the orientation preference of the receptive
     fields encoded in terms of colours, and demonstrating that the
     visual cortex performs an explicit
     expansion of the receptive field shapes over spatial image
     orientations.}
   \label{fig-koch-ori-map}
 \end{figure}
 
\begin{figure}[hbtp]
   \begin{center}
     \begin{tabular}{c}
       {\em\small First-order affine Gaussian derivative kernels\/} \\
       \includegraphics[width=0.35\textwidth]{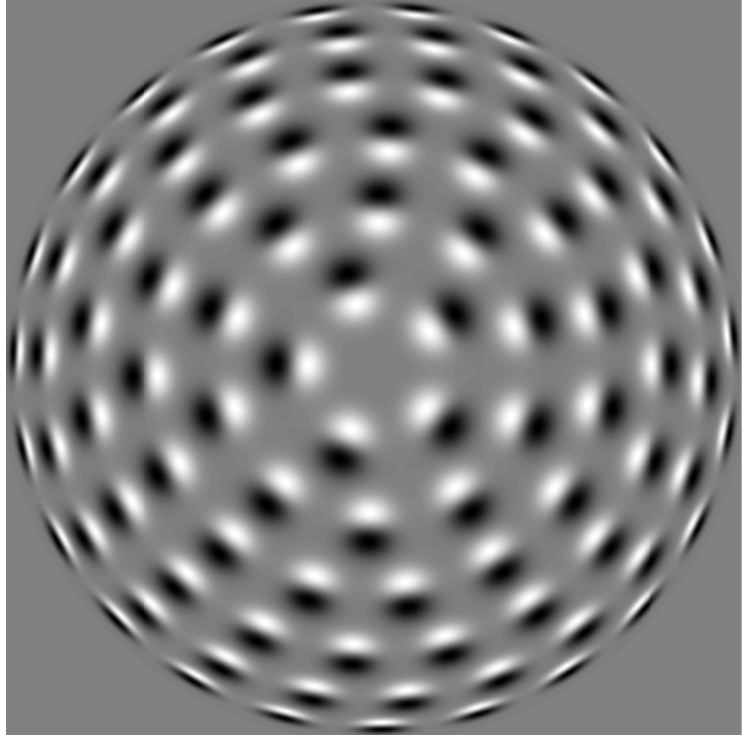} 
     \end{tabular}
   \end{center}
   \caption{Distribution of first-order affine Gaussian derivative
     kernels of the form (\ref{eq-spat-RF-model}) 
     for different spatial covariance matrices $\Sigma$, with
     their elements parameterized according to
     $C_{11} = \sigma_1^2 \, \cos^2 \varphi + \sigma_2^2 \, \sin^2 \varphi$,
     $C_{12} = C_{21} = (\sigma_1^2 - \sigma_2^2)  \cos \varphi  \, \sin \varphi$,
     and
     $C_{22} = \sigma_1^2  \, \sin^2 \varphi + \sigma_2^2  \, \cos^2 \varphi$,
     with the larger spatial scale parameter $\sigma_2$ in this
     illustration held constant,
     while the smaller scale parameter $\sigma_1$ varies as
     $\sigma_1 = \sigma_2/\kappa$, according to a distribution on a hemisphere.
     The spatial directional derivatives are, in turn, defined according to
     $\partial_{\varphi} =
     \cos \varphi \, \partial_{x_1} + \sin  \varphi \, \partial_{x_2}$.
     The possible additional variability of the scale parameters,
     beyond their ratio $\kappa$, is, however, not explicitly addressed in this
     paper. From a biological viewpoint, one could, indeed, possibly
     think that it might be easier to keep the smaller scale parameter
     $\sigma_1$ constant, and let the larger scale parameter $\sigma_2$
     increase towards the periphery, since a higher degree of
     orientation selectivity can then be achieved by just integrating over successively
     larger support regions in the image space. The important aspect of
     this illustration is rather that the
     eccentricity $\kappa$ increases from the most isotropic image
     position towards the periphery.
     With regard to the possible
     connection to the pinwheel structure in the primary visual
     cortex, the center in this figure would correspond to the center
     of the pinwheel, whereas the periphery would correspond to the
     boundaries of the part of the visual cortex that is closest to
     the center of one particular pinwheel.
     (Reprinted from Lindeberg (\citeyear{Lin23-FrontCompNeuroSci})
     (OpenAccess).)}
   \label{fig-aff-gauss-1der-distr}
\end{figure}

If additionally, reconstructions of the receptive field shapes could be
performed for the receptive fields probed during such
a systematic investigation of the difference in
response characteristics with the distance from the pinwheel centers,
and if the receptive fields could additionally be reasonably well modelled according
to the generalized Gaussian model for receptive fields studied and
used in this paper. Then, it would be interesting to investigate if the
shapes of the affine Gaussian components of these receptive fields would
span a larger part of the affine group, than the span over mere image
orientations, as already established in the orientation maps of the visual
cortex, as characterized by Bonhoeffer and Grinvald (\citeyear{BonGri91-Nature}),
Blasdel (\citeyear{Bla92-JNeuroSci}) and others, see
Figure~\ref{fig-koch-ori-map} for an illustration.
A working hypothesis in the paper concerns to investigate whether
the primary visual cortex could {\em additionally\/} perform an expansion over the
eccentricity or the elongation of the spatial components of the receptive fields.

If we would lay out the shapes of affine Gaussian receptive fields
according to the shapes of their underlying spatial covariance matrices $\Sigma$,
then we would for a fixed value of their size (the spatial scale parameter)
obtain a distribution of the form shown in Figure~\ref{fig-aff-gauss-1der-distr}.
That directional distribution is, however, in a certain aspect
redundant, since opposite orientations on the unit circle are
represented by two explicit copies, where the corresponding receptive
fields are either equal, for receptive fields corresponding to spatial
directional derivatives of even order, or of opposite sign for derivatives of even order.
Could it be established that the receptive fields shapes, if expanded
over a variability over eccentricity or elongation, for animals
that have a clear pinwheel structure, have a spatial
distribution that can somehow be related to such an idealized
distribution, if we collapse opposite image orientations to the same
image orientation, by {\em e.g.\/} a double-angle mapping
$\varphi \mapsto 2 \varphi$?

Notably the variability of the spatial covariance matrices in the
affine Gaussian derivative model comprises a variability over two
spatial scale parameters $\sigma_1$ and $\sigma_2$, while the
theoretical analysis of the orientation selectivity properties studied
in this article has mainly concerned their ratio
$\kappa = \sigma_2/\sigma_1$. Hence, the illustration in
Figure~\ref{fig-aff-gauss-1der-distr} should not be taken as a literal
prediction, even if reduced by a double-angle representation.
In Figure~\ref{fig-aff-gauss-1der-distr}, the larger scale parameter
$\sigma_1$ is held constant, for convenience of graphical
illustration, as obtained by mapping the uniformly sized receptive fields from a uniform
distribution on the hemisphere. More generally, one could also
conceive other distributions as possible, such as
instead keeping the smaller eigenvalue $\sigma_2$ constant
from the center towards the periphery.

To conclude, we propose to state the following testable hypotheses for
biological experiments:

\smallskip

\noindent
{\bf Hypothesis 1 (Variability in eccentricity)}
Let $\sigma_{\varphi}$ and $\sigma_{\bot\varphi}$ be the characteristic lengths in the
preferred directions of an orientation selective simple cell in the
primary visual cortex. Then, over a population of such simple cells,
there is a substantial variability in their eccentricity ratio
$\epsilon = \sigma_{\varphi}/\sigma_{\bot\varphi}$.

\smallskip

\noindent
{\bf Hypothesis 2 (Variability in eccentricity coupled to
  orientational homogeneity)}
Assuming that Hypothesis~1 holds, let $\epsilon$ denote the eccentricity
of a simple cell in the primary visual cortex, and let $H$ be a
measure of the homogeneity in the orientation preference of its
surrounding neurons. Then, over a population of simple cells, there is a
systematic connection between $\epsilon$ and $H$.

\smallskip

\noindent
{\bf Hypothesis 3 (Variability in eccentricity coupled to the pinwheel
  structure)}
Assuming that Hypothesis~1 holds, let $\epsilon$ denote the eccentricity
of a simple cell in the primary visual cortex. Then, over a population
of simple cells, there is a
systematic connection between $\epsilon$ and the distance from the
nearest pinwheel center.

\smallskip

If Hypothesis~3 would hold, then we could also sharpen this hypothesis further as:

\smallskip

\noindent
{\bf Hypothesis 4 (Increase in elongation with increasing distance from
  the centers of the pinwheels)}
Assuming that Hypothesis~3 holds, let $\epsilon$ denote the eccentricity
measure of a simple cell in the primary visual cortex defined such that 
$\epsilon = 1$ if the characteristic lengths of the spatial
receptive fields are equal, and tending towards zero as the
characteristic lengths differ more and more. Then, over a population
of simple cells, the eccentricity
measure decreases from the center of the pinwheel towards the
periphery.

\smallskip

Note that the latter explicit hypotheses have been expressed on a
general form, of not explicitly assuming that the biological receptive
fields can be well modelled according to the generalized Gaussian
derivative model for receptive fields. The essential factor in the
definitions is only that it should be possible to define estimates of the
characteristic lengths $\sigma_{\varphi}$ and $\sigma_{\bot\varphi}$,
so as to be able to define a measure of the eccentricity $\epsilon$.

\smallskip

If either Hypothesis~2 or Hypotesis~3 would hold, then we could also explicitly state the
following hypothesis:

\smallskip

\noindent
{\bf Hypothesis 5 (Pinwheel structure more structured than a mere
  expansion over spatial orientations)}
The pinwheel structure comprises an, at least, two-dimensional
variability of receptive field shapes, beyond an expansion over
spatial orientations, also an expansion over the eccentricity of the receptive fields
in the primary visual cortex.

\smallskip

For simplicity, we have above expressed these hypotheses for the case
of simple cells, for which it is easiest to define the measures
$\sigma_{\varphi}$ and $\sigma_{\bot\varphi}$ of the characteristic
lengths, because of the linearity of the receptive fields.
Provided that corresponding measures of characteristic lengths could
also be in a sufficiently well-established way be defined also for
non-linear complex cells, corresponding explicit hypotheses could also
be formulated for complex cells.

It should finally be stressed that, in this treatment, we have not considered
the binocular aspects of the pinwheel structure. In Hypothesis~5, the
variability of the pinwheel structure over contributions from the left
and the right eyes should therefore not be counted as a property to
contribute to the terminology ``more structured''.

\subsection{Quantitative measurements for detailed characterization}
\label{sec-quant-meas}

To further characterize possible relationships between the
orientational selectivity, receptive field eccentricity, orientational
homogeneity, and the location of the neuron in relation to the
pinwheel structure in the primary visual cortex, we would also propose
to characterize the possible relationships between these entities in
terms of:

\smallskip

\noindent
{\bf Quantitative measurement 1: (Relationship between orientational selectivity and
  receptive field eccentricity)}
Graph or scatter diagram showing how a quantitative measure of 
orientational selectivity is related to a quantitative measure of
receptive field eccentricity, accumulated over a sufficiently large
population of neurons.

\smallskip

\noindent
{\bf Quantitative measurement 2: (Relationship between orientational homogeneity and
  receptive field eccentricity)}
Graph or scatter diagram showing how a quantitative measure of 
orientational homogeneity is related to a quantitative measure
receptive field eccentricity, accumulated over a sufficiently large
population of neurons.

\smallskip

\noindent
{\bf Quantitative measurement 3: (Relationship between receptive field eccentricity
  and the pinwheel structure)}
Graph or scatter diagram showing how a quantitative measure
of receptive field eccentricity
depends on the distance to the nearest pinwheel center,
accumulated over a sufficiently large
population of neurons.

\smallskip

\noindent
{\bf Quantitative measurement 4: (Relationship between receptive field eccentricity
  and the pinwheel structure)}
Two-dimensional map showing how a quantitative measure of receptive field eccentricity
relates to a two-dimensional map of the orientation preference over
the same region in the primary visual cortex,
with the center of the pinwheel structure explicitly marked, again accumulated over a sufficiently large
population of neurons.

\smallskip

If the above theoretically motivated biological hypotheses could be
investigated experimentally, and if the above quantitative
measurements of receptive field characteristics could
be performed. Then, it could be judged if the prediction from the presented
theoretical analysis about a systematic variability in receptive field
eccentricity, with a possible relationship to the pinwheel structure,
could be either experimentally supported or rejected. In a corresponding
manner, such a judgement could also answer if the receptive fields in
the primary visual cortex could be regarded as spanning
a larger part of the affine group, than
an expansion over mere rotations in the image domain.

\section{Relations to other types of models of visual receptive fields}

Concerning the analysis presented in the previous section,
it should be emphasized that it constitutes a theoretical modelling
step at a {\em functional\/} mathematical level. The complementary theoretical
explanation in terms of a partial affine covariance property regarding
the degree of freedom corresponding to non-uniform scaling
transformation does furthermore operate at a coarse level of
abstraction, regarding theoretically desirable properties of
an idealized vision system.

Hence, this theoretical explanation should not be regarded as
in conflict with other possible theoretical explanations in
terms of learning of spatial receptive fields by sparse coding,
which may also lead to learned
receptive fields with different degrees of elongation
(Olshausen and Field \citeyear{OlsFie96-Nature},
\citeyear{OlsFie97-VR},
Rehn and Martin \citeyear{RehSom07-JCompNeuroSci},
Zylberberg {\em et al.\/} \citeyear{ZylMurDeW11-PlosCompBio},
King {\em et al.\/} \citeyear{KinZylDeW13-JNeuroSci}).
Similarly, our conceptual explanation is neither in conflict
with a more fine-grained computational explanation
of varying receptive field shapes in terms of different
``On'' or ``Off'' responses within the receptive fields,
as proposed by Martinez {\em et al.\/}
(\citeyear{MarWanReiPilAloSomHir05-NatNeuroSci}) and
neurophysiologically investigated by
Kremkow {\em et al.\/}\ (\citeyear{KreJinWanAlo16-Nature})
and Jansen  {\em et al.\/}
(\citeyear{JanJinLiLasKreBerSwaZaiAlo19-CerebrCort}).

Instead, our proposed explanation is that the notion of affine
covariance or more generally covariance to geometric image
transformations may be an essential factor in the development
of biological vision systems. As natural images are formed in the
retinas of visual observers, the statistical properties of the image
data that reach the visual sensor will be strongly influenced by
the structure of the natural image transformations,
when 3-D objects are projected to 2-D image data from
different viewing positions, viewing directions and relative motions.

The result of learning receptive fields in a biological vision agent,
should thereby be strongly influenced by the structure of the
natural image transformations. For a biological creature, who
depends strongly on the visual perception for its survival,
it does therefore seem plausible that there could be an
evolutionary pressure for the biological organism to develop
its vision system to be well adapted to the influence of the
natural image transformations that arise when observing
a dynamic world.
Beyond the treatment of whether biological
vision has actually developed a variability in the eccentricity
of the receptive fields, our theory offers a theoretical explanation
of this property, solidly grounded in a principled formal theory of visual
receptive fields.

Thus, one could expect that when specific learning approaches,
such as the sparse coding mechanisms considered by
Olshausen and Field (\citeyear{OlsFie96-Nature},
\citeyear{OlsFie97-VR}),
Rehn and Martin (\citeyear{RehSom07-JCompNeuroSci}),
Zylberberg {\em et al.\/} (\citeyear{ZylMurDeW11-PlosCompBio}) and
King {\em et al.\/} (\citeyear{KinZylDeW13-JNeuroSci}), are
exposed to natural image data that contain the variabilities
in image structures generated by geometric image transformations,
it would be natural for such learning approaches to
lead to learned receptive fields with different eccentricities.
Specifically, one could then also conceive that different spatial shapes
of such receptive fields to be biologically implemented in
terms of different contributions form ``On'' and ``Off'' subregions,
as considered by
Martinez {\em et al.\/}
(\citeyear{MarWanReiPilAloSomHir05-NatNeuroSci}),
Kremkow {\em et al.\/}\ (\citeyear{KreJinWanAlo16-Nature})
and Jansen  {\em et al.\/}
(\citeyear{JanJinLiLasKreBerSwaZaiAlo19-CerebrCort}).

In this respect, the presented theory in this paper thus offers a
theoretical explanation of these phenomena
at a more general and functional level of abstraction, by
assuming an idealized vision system that adapts its
processing to the inherent structures of
natural image structures as influenced by natural image
transformations. The property that the receptive fields in the primary
visual cortex should have a variability in their
eccentricity is then a direct consequence of this general
assumption, {\em via\/} the
assumption of affine covariance.
This could then also conceptually explain
other possible explanations in terms of more explicit learning
mechanisms or specific neurophysiological mechanisms.

In Appendices~\ref{sec-rel-gabor-model}
and~\ref{sec-rel-fine-grained-models}
we additionally:
(i)~give relations to a corresponding analysis based on Gabor
models of the visual receptive fields, and
(ii)~describe relations to more detailed and fine-grained models
of the primary visual cortex.

\section{Summary and discussion}
\label{sec-summ-disc}

We have compared results from
theoretical analysis of the orientation selectivity
properties of the affine Gaussian derivative model
(Section~\ref{sec-conn-ori-sel-elong}) with
experimental results by
Nauhaus {\em et al.\/}\ (\citeyear{NauBenCarRin09-Neuron})
on broadly {\em vs.\/} sharply tuned visual neurons
(Figure~\ref{fig-nauhaus-ori-tun-meas}) and
by Goris {\em et al.\/}\ (\citeyear{GorSimMov15-Neuron})
on rather uniform distributions of the resultant values from orientation
selectivity curves (Figure~\ref{fig-goris-orient-select-stat}).
Thereby, we have found potential support for one of the
dimensions of variability in a biological hypothesis formulated
in Lindeberg (\citeyear{Lin23-FrontCompNeuroSci}),
stating that the family
of receptive field shapes ought to span the degrees of freedom in the
natural geometric image transformations.
This potential support rests on the assumption, that 
it should be unlikely
for the population of receptive fields 
to show strong variability in orientation selectivity,
without also showing similar variability in eccentricity
or elongation.

Without explicitly relying on expressing such an explicit assumption,
regarding whether the visual receptive fields in the primary visual
cortex could be well modelled by affine Gaussian derivative based
receptive fields, we can, however, firmly state that the biological
measurements performed by Nauhaus {\em et al.\/}\
(\citeyear{NauBenCarRin09-Neuron}) and by
Goris {\em et al.\/}  (\citeyear{GorSimMov15-Neuron})
are, in combination with the
theoretical results summarized in Section~\ref{sec-conn-ori-sel-elong},
consistent with the hypothesis that the receptive fields should
span a variability in the eccentricity of the receptive fields, and
more widely consistent with the hypothesis about affine covariant receptive
fields.


Furthermore, from the results presented in
Section~\ref{sec-interpret-goris-exps-3-4}, that the orientation
selectivity histograms accumulated by
Goris et al.\ (\citeyear{GorSimMov15-Neuron}) appear to be better
explained by a population of receptive fields up to order 4 than by
receptive fields up to order 2, one may speculate if it would be
appropriate to put additional focus of neurophysiological experiments
on exploring the presence of visual receptive fields of higher order,
compared to the receptive fields of lower order more commonly
reported in the neurophysiological literature.

If we apply a similar type of assumption-based logical reasoning to the pinwheel
structure in the primary visual cortex, then such a reasoning, based
on the results by Nauhaus {\em et al.\/}\
(\citeyear{NauBenCarRin09-Neuron})
and by Wilson {\em et al.\/} (\citeyear{WilWhiSchFit16-NatNeuroSci}),
that the orientation selectivity
appears to vary strongly from the centers of the pinwheels towards the
periphery.
Then, this implies that the pinwheel structure in the visual cortex would, beyond
an explicit expansion over image orientations, also comprise an explicit
expansion over the the degree of elongation of the
receptive fields. Based on these predictions, we propose to consider
explicit dependencies on a variability in the
eccentricity of the receptive fields,
when modelling the pinwheel structure in the primary visual cortex.

Strictly, and formally, the results from such logical inference could,
however, only be regarded as theoretical predictions, to generate
explicit hypothesis concerning the distribution of receptive field
characteristics in these respects. To raise the question of
determining if these theoretical predictions would firmly hold in reality, 
we propose that the testable explicit biological hypotheses formulated in
Section~\ref{sec-biol-hypo} could be used to, in neurophysiological
experiments, either verify or reject the overall hypothesis, concerning possible
variabilities in the eccentricity of the receptive fields in the
primary visual cortex of higher mammals. These predictions
could also be used to explore hypotheses about possible
connections between such variabilities in the eccentricity or the elongation and other
receptive field characteristics, in particular in relation to the
pinwheel structure in the primary visual cortex of higher mammals.

Notably, the overall analysis in the paper is carried based on
a framework that reflects functional properties of an idealized vision
  system, in terms of covariance under geometric image transformations,
  and using mathematical analysis as the main tool.
  Thereby, the analysis in the paper is performed without any extensive numerical
  simulations of explicit neuron models, which would otherwise have
  to contain many parameters,
  some of which commonly are unobservable and would have to be estimated
  indirectly from data.
  
Concerning possible limitations in the hypothetical reasoning stages used for
possible logical inference and for formulating the explicit biological hypotheses
above, the logical reasoning based on connections between the
eccentricity
and the degree of elongation of the receptive fields depend
on explicitly stated assumptions regarding whether the
biological receptive fields
could be reasonably well modelled by affine Gaussian derivative based
receptive fields. To be able to draw possible further conclusions,
the possible validity of those hypothetical logical reasoning
stages could, however, break down, if there would be other external
factors, not covered by the theoretical model, that could also
strongly influence the orientation selectivity of the receptive fields.
The possible applicability of the hypothetical logical reasoning
stages above thus, ultimately, depends on the possible agreement between
the model and biological data, and can only be taken further by
performing complementary neurophysiological
experiments, to ultimately judge if the theoretically based
predictions, stated in Section~\ref{sec-biol-hypo},
would be applicable to actual biological neurons.

Furthermore, while the treatment in this paper has been concerned with
species that have orientation maps and a pinwheel structure, a very
interesting follow-up question would then also concern if
corresponding results would extend to species that do not have
structured orientation maps or a pinwheel structure, and then
specifically which such species.
For example, Niell and Stryker (\citeyear{NieStr08-JNeuroSci})
have shown that the mouse visual cortex has developed
mechanisms of orientation selectivity, although not having
pinwheel structures.
For analyzing that topic in detail, further results from biological
experiments and on other species would, however, be necessary,
why we leave that topic to future work.

\section*{Appendix}

\appendix

\section{Relations to a corresponding analysis based on Gabor models}
\label{sec-rel-gabor-model}

In view of the presented analysis of a predicted expansion of
receptive field shapes over the degree of elongation of the receptive
fields, one may naturally ask if a corresponding type of analysis
could also be carried out for the Gabor model of visual receptive
fields. For such a purpose, let us consider an affine Gabor model of
the form
\begin{align}
  \begin{split}
    \label{eq-aff-gabor-mod-even}
    T_{\even}(x_1, x_2;\; \sigma_1, \sigma_2, \nu)
    & = \frac{1}{2 \pi \sigma_1 \sigma_2} \,
            e^{- x_1^2/2 \sigma_1^2 - x_2^2/2 \sigma_2^2}
            \cos(\nu \, x_1),
    \end{split}\\
  \begin{split}
    \label{eq-aff-gabor-mod-odd}    
    T_{\odd}(x_1, x_2;\; \sigma_1, \sigma_2, \nu)
    & = \frac{1}{2 \pi \sigma_1 \sigma_2} \,
            e^{- x_1^2/2 \sigma_1^2 - x_2^2/2 \sigma_2^2}
            \sin(\nu \, x_1),
    \end{split}
\end{align}
where $\sigma_1$ and $\sigma_2$ are the spatial scale parameters and
$\nu$ the angular frequency. For simplicity, we have here oriented the
affine Gabor model to the image orientation $\varphi = 0$.

In Lindeberg (\citeyear{Lin25-JCompNeurSci-orisel}) Section~5, it is shown
that if we couple the spatial scale parameters $\sigma_1$ and $\sigma_2$
in that model according to $\sigma_2 = \kappa \, \sigma_1$ for $\kappa > 1$,
then larger values of the scale parameter $\kappa$ will lead to more narrow
orientation selectivity properties. There is, however, also another
degree of freedom in the affine Gabor model, implying that variations in the angular
frequency $\nu$ may also strongly affect the orientation selectivity
properties of the receptive fields.

Specifically, comparing the explicit expressions
for the orientation selectivity curves derived for the affine Gabor model
in Lindeberg (\citeyear{Lin25-JCompNeurSci-orisel})
\begin{align}
  \begin{split}
    A_{\even}(\theta)
    = \frac{1}{2} \left(e^{2 \nu ^2 \sigma_1^2 \cos \theta}+1 \right) 
        e^{- \nu ^2 \sigma_1^2 \cos ^2 \left(\frac{\theta }{2}\right)
        (\kappa^2-1)(1 -  \cos \theta)},
  \end{split}\\
  \begin{split}
    A_{\odd}(\theta)
    = \frac{1}{2} \left( e^{2 \nu ^2 \sigma_1^2 \cos \theta}-1 \right)
        e^{- \nu ^2 \sigma_1^2 \cos ^2 \left(\frac{\theta }{2}\right)
        (\kappa^2-1)(1 -  \cos \theta)}.
  \end{split}
\end{align}
to the explicit expressions for the orientation selectivity curves for
the generalized Gaussian derivative model for visual receptive fields,
summarized to the form 
(\ref{eq-ori-sel-curves-gen-gauss-der-model-general})
\begin{equation}
  \label{eq-ori-sel-curves-gen-gauss-der-model-general-again}
  r_{\lambda}(\theta)
  = \left(
         \frac{\left| \cos \theta \right|}
         {\sqrt{\cos ^2 \theta + \kappa ^2 \sin ^2\theta}}
         \right)^{\lambda}
\end{equation}
for a few discrete values of $\lambda$,     
we see that the relationship between the degree of elongation of a
receptive field and the orientation selectivity
properties of the receptive field is rather complex for the affine Gabor
model, while it is very direct for the generalized
Gaussian derivative model for visual receptive fields.

Thus, if we assume that the population of the visual neurons would be
described by an affine Gabor model, and if we observe variations in
the orientation selectivity properties of the neurons, in terms of
``more narrow'' or ``less narrow'', then it is, in fact,
{\em not possible\/} to logically infer that those variations in the
orientation selectivity would
necessarily have to be due to variations in the scale parameter ratio $\kappa$, since
those variations in the orientation selectivity properties could also
be due to variations in the other degree of freedom, determined by the
interactions between the angular frequency $\nu$, the spatial scale
parameter $\sigma_1$ and the scale parameter ratio $\kappa$.

Regarding the use of a Gabor model for the visual receptive fields, it
should also be stressed that the Gabor model is a purely spatial
model, while the biological receptive fields in the primary visual
cortex do also have strong temporal dependencies. A further
advantage of the generalized Gaussian derivative model for visual
receptive fields in that context, is that it also comprises
theoretically principled models for joint spatio-temporal receptive
fields.

Additionally, we have in
Lindeberg (\citeyear{Lin25-JCompNeurSci-orisel}) shown that the
orientation selectivity curves for the joint spatio-temporal receptive
field models
(\ref{eq-spat-temp-RF-model-der-norm-caus}) and
(\ref{eq-quasi-quad-dir-vel-adapt-spat-temp})
up to order 2
lead to similar orientation selectivity curves
(\ref{eq-ori-sel-r-simple1})--(\ref{eq-ori-sel-r-complex})
as for the corresponding purely spatial models
(\ref{eq-spat-RF-model}) and (\ref{eq-quasi-quad-dir}).
In this respect, the analysis based on the generalized Gaussian
derivative model for visual receptive fields has a much wider domain of
validity, compared to a corresponding analysis based on the affine
Gabor model for visual receptive fields, since its results hold also
for the more realistic domain of joint spatio-temporal receptive fields.

\section{Relations to more detailed and fine-grained models of the primary visual cortex}
\label{sec-rel-fine-grained-models}

Compared to more more detailed and fine-grained quantitative models of the primary
visual cortex
(Troyer {\em et al.\/} \citeyear{TroKruPriMil98-JNeuroSci},
Zhu {\em et al.\/} \citeyear{ZhuSheSha09-JCompNeuroSci},
Shu {\em et al.\/} \citeyear{ShuGaCheLiu15-PONE},
Schmidt {\em et al.\/} \citeyear{SchBakHilDieAlb18-BrainStructFunc},
Einevoll {\em et al.\/} \citeyear{EinDesDieGruJirKamMigNesPleSch19-Neuron},
Billeh {\em et al.\/} \citeyear{BilCaiGraDaiIyeGouAbbJiaSieOlsKocMihArk20-Neuron},
Chariker {\em et al.\/} \citeyear{ChaShaHawYou22-JNeuroSci},
Antol{\'\i}k {\em et al.\/}  \citeyear{AntCagRozMonFreDav24-PLOSCompBiol}),
a major conceptual difference of the presented highly idealized and
mathematically principled model, is that the predictions and the
results can be obtained with just straightforward theoretical
modelling and analysis, and essentially, thus, be obtained in closed
form, as opposed to the results from numerical simulations of comparably much more
complex models, then implying a need for both (i)~further assumptions,
explicit or implicit, for formulating the actual models, as well as for
(ii)~setting the values of the possibly rather large number of
parameters in such more complex models.
Specifically, it is most often the case that some parameters are not experimentally measured and therefore need to be obtained by parameter estimation (parameter tuning) according to functional criteria, thereby necessitating these functional criteria to be formulated, further complicating the use of complex models.

Whereas one could possibly argue that the particle simulation approach,
used for accumulating the receptive field histograms in
Section~\ref{sec-interpret-goris-exps-1-2}
and~\ref{sec-interpret-goris-exps-3-4},
from explicitly generated samples from the distributions over the scale
parameter ratio $\kappa$,
could also be regarded as constituting numerical simulations with a
certain numerical uncertainty, due to sampling issues of
the underlying distributions,
it would, in principle, also be possible to
instead obtain more accurate estimates of the corresponding receptive
field histograms without using any use of such sampled distributions.
Instead, one can numerically solve the equations that determine
the boundaries of the
quantization bins in the distributions of the receptive fields, which,
in turn, form the histograms as integrals of the underlying
distributions over the resulting bin regions in the parameter space.
Such an approach would then lead to results
without any dependence on sampling issues regarding the model. In the experimental
results reported in this paper, such possible sampling issues would
nevertheless be expected to be very small, due
to the rather large number of samples underlying the accumulation of
the sampled histograms. Furthermore, the focus of this paper is on
qualitative properties with regard to the distributions of receptive fields over
different degrees of elongation.

While large scale models could be very valuable, if they could be
properly aligned with biological measurements, more detailed and
fine-grained models of the primary visual cortex, do on the other hand
require numerical calculations to be evaluated, and do
therefore rely on explicit numerical simulations for getting the
results. Hence, a large number of larger-scale simulations would be
needed to get corresponding results as presented in this
paper. Then, also those results would furthermore depend on the assumptions used
for constructing those models as well as on the choices of the
complementary parameters in those substantially more complex models, that
they are based on. To interpret such corresponding results, regarding
the possible validity of the hypothesis of an expansion of receptive
field shapes over the degree of elongation, rather exhaustive
searches in the resulting parameter spaces may therefore be needed, to
determine the regions in the rather high-dimensional parameter spaces,
where the stated predictions would hold, and then also delimit the
boundaries in the parameter space of such valid regions. A particular
critical aspect of such an analysis, aimed at exploring whether the
receptive fields would span a variability over the degree of
elongation or not,  would, however, concern if the
architecture for the fine-grained model would, based on either
implicit or explicit assumptions, either support or violate affine covariance.

If we consider the perspective of judging whether the receptive field shapes
could be regarded as spanning a substantial variability over the degree
of elongation. Then, the proposed model constitutes an as much simplified
and idealized theoretical model as possible, while
still respecting the covariance properties of the receptive fields
under geometric image transformations.
The model is, thus, also
essentially parameter free, beyond the bounds on the scale parameter
ratio $\kappa$, that were handled by symmetric
intervals of the form $[1/\kappa_{\max}, \kappa_{max}]$ for the here,
for purposes of illustration only,
arbitrary choice of $\kappa_{\max} = 8$, and the above, for simplicity,
in the absence of further information, assumed equal numbers of
receptive fields corresponding to the different orders of spatial differentiation,
in the composed
histograms accumulated over receptive fields corresponding to multiple orders of
spatial differentiation.
Additionally, the underlying model for the visual receptive
fields, which the presented work is based on, is theoretically very
principled, by being derived in a purely axiomatic manner, and then
after the axiomatic mathematical derivation {\em a posteriori\/} compared to
actual neurophysiological recordings of biological receptive fields.

The model, that the presented analysis is based on, is also a purely
functional model for feedforward computations in V1, and thereby
decoupled from the more complex workings of the neuronal architecture
in V1. From this perspective, the analysis shows that the variability
in the degree of elongation of the receptive fields ought to
constitute a very basic property of the receptive fields in V1,
that to a first order of approximation would not necessarily explicitly depend
upon such more complex complementary mechanisms. 

A very interesting topic for future studies could nevertheless be to
complement the presented theory
with the influence of more complex neural mechanisms, such
intra-V1-coupling mechanisms, the top-down influence in the visual
cortex and inhibitory projections.
Another interesting approach would be to integrate mechanisms for
affine covariance into more fine-grained V1 models.
Such more fine-grained modelling would then have the possibility to
model detailed substructures in the reported
orientation selectivity properties of the visual neurons, that are not
comprised within the realm of our maximally idealized model.
For example, Liu {\em et al.\/}
(\citeyear{LiuHasLyo15-JPhys,LiuHasLyo17-NeuroPhot}) have
reported that the orientation tuning between the pinwheel and
domain neurons in primary visual cortex may additionally depend also on
the stimulus contrast and the stimulus size.%
\footnote{Furthermore, an additional dependency on stimulus size could also
  result from an expansion of receptive field shapes over the size of
  the receptive fields, as also predicted from the property of affine
  covariance, that underlies the normative theory of visual receptive
  fields, used as the theoretical foundation for this study.}
From a theoretical viewpoint, such complementary
dependencies on stimulus contrast could be explained by the
also non-linear dependencies of the receptive fields on the input
stimuli, which may then
lead to qualitatively different types of receptive fields from
different stimuli, for
example, based on mechanisms such as inhibitory tuning
(Li {\em et al.\/}\ \citeyear{LiMaLiIbrWanTao12-JNeuroSci})
or, more generally, that the contrast of the visual stimuli may modulate
the functional connectivity in the visual cortex
(Nauhaus {\em et al.\/}\ \citeyear{NauBusCarRin90-NatNeuroSci}),
see, however, also Alitto and Usrey (\citeyear{AliUsr04-JNeuroPhys})
and Nowak and Barone (\citeyear{NowBar09-PONE})
for complementary support regarding contrast invariant properties
of the orientation tuning of the visual neurons.

In the view of such more detailed and fine-grained modelling of the
neural architecture in V1,
it would, indeed, be highly interesting to complement the
receptive field models in our normative theory
with such complementary mechanisms, to adapt it to the
more fine-grained properties in the primary visual cortex.
Since the exploration of such
mechanisms would, however, require explicitly carrying out a substantial number of
numerical simulations of much larger and more complex V1 models, 
we leave such extensions to future work, with the focus of this
paper on purely theoretically based ways of reasoning based on our
maximally simplified idealized model.

A main advantage of the presented work, based on the underlying
normative theory for visual receptive fields, 
is therefore, that it, on the other
hand, shows that it is possible to perform substantial analysis, reasoning and
predictions based on a purely theoretical framework, which then enables much
more compact and efficient analysis, compared to more extensive
numerical simulations of more complex, detailed and fine-grained
models of the primary visual cortex. As has been demonstrated, the
results from this theoretical way of reasoning are also in very good
qualitative agreement with the results from previously reported biological experiments.

\section{Derivations of orientation selectivity properties for third-order and fourth-order simple cells}
\label{sec-ori-sel-order-3-4}

For purposes in the main article, we here extend the results
concerning the orientation selectivity curves for idealized models of
cells according to the
generalized Gaussian derivative model
for visual receptive fields in  (Lindeberg
\citeyear{Lin25-JCompNeurSci-orisel}) from first-order and second-order simple
cells to also comprise third-order and fourth-order simple cells.

For simplicity, we will here restrict ourselves to purely static
models of simple cells.

\subsection{Third-order simple cell}

Following the methodology in (Lindeberg
\citeyear{Lin25-JCompNeurSci-orisel})
underlying the results summarized in
Section~\ref{sec-ori-sel-curves-ideal-rf-models}
we will express an idealized model of a simple cell
with four lobes along the preferred orientation of the simple cell
as a third-order scale-normalized derivative of an affine Gaussian kernel
(according to
(\ref{eq-spat-RF-model})
for $m = 3$), and for
convenience of the calculations choose the preferred orientation as the
horizontal $x_1$-direction (for $\varphi = 0$) with spatial scale parameter $\sigma_1$ in
the horizontal $x_1$-direction and spatial scale parameter $\sigma_2$
in the vertical $x_2$-direction, and thus with a spatial covariance matrix
of the form $\Sigma_0 = \diag(\sigma_1^2, \sigma_2^2)$:
\begin{align}
  \begin{split}
    & T_{000,\norm}(x_1, x_2;\; \sigma_1, \sigma_2) =
  \end{split}\nonumber\\
  \begin{split}
     & = \frac{\sigma_1^3}{2 \pi \sigma_1 \sigma_2} \,
            \partial_{x_1 x_1 x_1}
              \left( e^{-x_1^2/2\sigma_1^2 - x_2^2/2 \sigma_2^2} \right)
  \end{split}\nonumber\\
  \begin{split}
    & = \frac{(3 \sigma_1^2 x - x^3)}{2 \pi \sigma_1^4 \sigma_2} \,
               e^{-x_1^2/2\sigma_1^2 - x_2^2/2 \sigma_2^2}.
  \end{split}
\end{align}
The corresponding receptive field response can then be expressed as, after solving the
convolution integral in Mathematica,
\begin{align}
   \begin{split}
     L_{000,\norm}(x_1, x_2;\; \sigma_1, \sigma_2) =
  \end{split}\nonumber\\
  \begin{split}
    & = \int_{\xi_1 = -\infty}^{\infty}  \int_{\xi_2 = -\infty}^{\infty}
             T_{000,\norm}(\xi_1, \xi_2;\; \sigma_1, \sigma_2)
  \end{split}\nonumber\\
  \begin{split}
    & \phantom{= = \int_{\xi_1 = -\infty}^{\infty}  \int_{\xi_2 = -\infty}^{\infty}}
             \times f(x_1 - \xi_1, x_2 - \xi_2) \, d \xi_1 \xi_2
  \end{split}\nonumber\\
  \begin{split}
    & = - \omega^3 \, \sigma_1^3 \cos^3 (\theta) \,
           e^{-\frac{1}{2} \omega^2 (\sigma_1^2 \cos^2 \theta + \sigma_2^2 \sin^2 \theta)}
  \end{split}\nonumber\\
  \begin{split}
    \label{eq-L000-pure-spat-anal}
    & \phantom{= =}
           \times \cos
             (
                \omega \cos (\theta) \, x_1 + \omega \sin (\theta) \, x_2 + \beta
             ),
   \end{split}         
\end{align}
{\em i.e.\/},\ it corresponds to cosine wave with amplitude
\begin{multline}
  A_{\varphi\varphi\varphi}(\theta, \omega;\; \sigma_1, \sigma_2) = \\
  = \omega^3 \, \sigma_1^3 \cos^3 (\theta) \,
      e^{-\frac{1}{2} \omega^2 (\sigma_1^2 \cos^2 \theta + \sigma_2^2 \sin^2 \theta)}.
\end{multline}
If we, for this modelling situation, assume that the spatial receptive
field is fixed, then it follows that the amplitude of the
response will strongly depend on the angular frequency $\omega$ of the
sine wave. Specifically, the magnitude of the response will first
increase with the angular frequency of the input stimulus, because of the
factor $\omega$. Then, it will decrease with scale because of the
strong exponential decrease with $\omega^2$.

Let us consider that a biological experiment to measure the
orientation selectivity properties of a visual neuron is performed
in such a way that the angular frequency of the input stimulus is
varied for each inclination angle $\theta$, and that then the result
for each value orientation $\theta$ of the stimulus is only reported
for the angular frequency $\hat{\omega}$ that leads to the maximum
response over all the image orientations.
Then, we can determine this value of $\hat{\omega}$ by
differentiating $A_{\varphi}(\theta, \omega;\; \sigma_1, \sigma_2)$ with
respect to $\omega$ and setting the derivative to zero, which gives:
\begin{equation}
  \hat{\omega}_{\varphi\varphi\varphi}
  = \frac{\sqrt{3}}{\sqrt{\sigma_1^2 \cos^2 \theta + \sigma_2^2 \sin^2 \theta}}.
\end{equation}
If we then insert this value into $A_{\varphi\varphi\varphi}(\theta, \omega;\; \sigma_1, \sigma_2)$,
and introduce a scale parameter ratio $\kappa$ such that
\begin{equation}
   \sigma_2 = \kappa \, \sigma_1,
 \end{equation}
which gives
\begin{equation}
  \label{eq-omega3-spat}
  \hat{\omega}_{\varphi\varphi\varphi}
  = \frac{\sqrt{3}}{\sigma_1 \sqrt{\cos^2 \theta + \kappa^2 \sin^2 \theta}}.
  \end{equation}
then this gives rise to an orientation selectivity curve of the form
\begin{equation}
    \label{eq-ori-sel-simple-3der}
  A_{\varphi\varphi\varphi,\max}(\theta, \; \kappa)
  = \frac{3 \sqrt{3} \left| \cos^3 \theta \right|}
     {e^{3/2} \left(\cos ^2 \theta + \kappa ^2 \sin ^2\theta \right)^{3/2}}.
\end{equation}
Notably, this amplitude measure is independent of the
spatial scale parameter $\sigma_1$ of the receptive field.
This property is a direct implication of the scale-invariant properties of
differential expressions in terms of
scale-normalized derivatives when using the specific value
$\gamma = 1$ for the scale normalization parameter.

\subsection{Fourth-order simple cell}

Let us next consider an idealized model of a simple cell with five
lobes along the main orientation of the receptive field, which
we model with as a fourth-order
scale-normalized derivative of an affine Gaussian kernel
(according to
(\ref{eq-spat-RF-model})
for $m = 4$), with its
preferred orientation again for convenience chosen as the
horizontal $x_1$-direction (for $\varphi = 0$), and
with a spatial scale parameter $\sigma_1$ in
the horizontal $x_1$-direction and a spatial scale parameter $\sigma_2$
in the vertical $x_2$-direction, and thus again with a spatial covariance matrix
of the form $\Sigma_0 = \diag(\sigma_1^2, \sigma_2^2)$:
\begin{align}
  \begin{split}
    & T_{0000,\norm}(x_1, x_2;\; \sigma_1, \sigma_2) =
  \end{split}\nonumber\\
  \begin{split}
     & = \frac{\sigma_1^4}{2 \pi \sigma_1 \sigma_2} \,
            \partial_{x_1 x_1 x_1 x_1}
              \left( e^{-x_1^2/2\sigma_1^2 - x_2^2/2 \sigma_2^2} \right)
  \end{split}\nonumber\\
  \begin{split}
    & = \frac{(3 \sigma_1^4 -6 \sigma_1^2  x^2+x^4)}{2 \pi \sigma_1^5 \sigma_2} \,
               e^{-x_1^2/2\sigma_1^2 - x_2^2/2 \sigma_2^2}.
  \end{split}
\end{align}
After solving the convolution integral in Mathematica,
the corresponding receptive field response is then of the form
\begin{align}
   \begin{split}
     L_{0000,\norm}(x_1, x_2;\; \sigma_1, \sigma_2) =
  \end{split}\nonumber\\
  \begin{split}
    & = \int_{\xi_1 = -\infty}^{\infty}  \int_{\xi_2 = -\infty}^{\infty}
             T_{0000,\norm}(\xi_1, \xi_2;\; \sigma_1, \sigma_2)
  \end{split}\nonumber\\
  \begin{split}
    & \phantom{= = \int_{\xi_1 = -\infty}^{\infty}  \int_{\xi_2 = -\infty}^{\infty}}
             \times f(x_1 - \xi_1, x_2 - \xi_2) \, d \xi_1 \xi_2
  \end{split}\nonumber\\
  \begin{split}
    & = \omega^4 \, \sigma_1^4 \cos^4 (\theta) \,
           e^{-\frac{1}{2} \omega^2 (\sigma_1^2 \cos^2 \theta + \sigma_2^2 \sin^2 \theta)}
  \end{split}\nonumber\\
  \begin{split}
    \label{eq-L0000-pure-spat-anal}
    & \phantom{= =}
           \times \sin
             (
                \omega \cos (\theta) \, x_1 + \omega \sin (\theta) \, x_2 + \beta
             ),
   \end{split}         
\end{align}
{\em i.e.\/},\ a sine wave with amplitude
\begin{multline}
  A_{\varphi\varphi\varphi\varphi}(\theta, \omega;\; \sigma_1,
  \sigma_2) = \\
  = \omega^4 \, \sigma_1^4 \cos^4 (\theta) \,
      e^{-\frac{1}{2} \omega^2 (\sigma_1^2 \cos^2 \theta + \sigma_2^2 \sin^2 \theta)}.
\end{multline}
As for the previous idealized receptive field model, this
expression also first increases and then increases with
the angular frequency $\omega$.
Again selecting the value of $\hat{\omega}$ at which the amplitude
assumes its maximum over $\omega$ gives
\begin{equation}
  \label{eq-omega4-spat}  
  \hat{\omega}_{\varphi\varphi\varphi\varphi}
  = \frac{2}{\sigma_1 \sqrt{\cos^2 \theta + \kappa^2 \sin^2 \theta}},
\end{equation}
which implies that the maximum amplitude over spatial scales as a
function of the inclination angle $\theta$ and the scale parameter
ratio $\kappa$ can be written
\begin{equation}
  \label{eq-ori-sel-simple-4der}
  A_{\varphi\varphi\varphi\varphi,\max}(\theta;\; \kappa)
  = \frac{16 \cos^4 \theta}
             {e^2 \left( \cos ^2 \theta + \kappa ^2 \sin ^2\theta \right)^2}. 
\end{equation}

\subsection{Resulting orientation selectivity curves}
\label{sec-ori-sel-curves-3-4}

If we additionally normalize the orientation selectivity curves
(\ref{eq-ori-sel-simple-3der}) and (\ref{eq-ori-sel-simple-4der}) to
have their maximum value equal to one for the preferred
orientation $\theta = 0$, we then obtain normalized
orientation selectivity curves of the forms
\begin{align}
  \begin{split}
    \label{eq-ori-sel-r-simple3-app}
    r_{\simple,3}(\theta)
    & = \frac{\left| \cos \theta \right|^3}
                   {(\cos ^2 \theta + \kappa ^2 \sin ^2\theta)^{3/2}},
  \end{split}\\
  \begin{split}
    \label{eq-ori-sel-r-simple4-app}    
    r_{\simple,4}(\theta)
    & = \frac{\cos^4 \theta}
                   {(\cos ^2 \theta + \kappa ^2 \sin ^2\theta)^2},
  \end{split}
\end{align}
and with examples of graphs of these curves, for a few values of the
scale ratio parameter $\kappa$, shown in 
in the bottom row of Figure~\ref{fig-ori-sel}.

\bibliographystyle{abbrvnat}

{\footnotesize
\bibliography{defs,tlmac}
}

\end{document}